\documentclass{aa}
\usepackage{graphics}
\usepackage{txfonts}
\usepackage{natbib}

\begin{document}

\title
{
The Variable X-ray Spectrum of Markarian 766 - II.
Time-Resolved Spectroscopy. 
\thanks{
Based on observations obtained with {{\it XMM-Newton}}, 
an ESA science mission with instruments and contributions 
directly funded by ESA Member States and NASA, also on 
observations obtained with {\it Suzaku}, a collaboration between 
ISAS/JAXA and NASA/GSFC, MIT
}
}
\subtitle{}

\titlerunning{Time-Resolved Spectroscopy of Mrk\,766}

\author
{T.\ J.\ Turner\inst{1,2} \and 
 L.\ Miller\inst{3} \and 
J.\ N.\ Reeves\inst{4} \and 
S.\ B.\ Kraemer\inst{2,5}
}

\authorrunning{T.J. \ Turner et al.\ }

\institute{Dept. of Physics, University 
of Maryland Baltimore County, 1000 Hilltop Circle, Baltimore, MD 21250, U.S.A.
\and  
Astrophysics Science Division,   
NASA/GSFC, Greenbelt, MD 20771, U.S.A.
\and 
Dept. of Physics, University of Oxford, 
Denys Wilkinson Building, Keble Road, Oxford OX1 3RH, U.K.
\and
Astrophysics Group, School of Physical and Geographical Sciences, Keele 
University, Keele, Staffordshire ST5 5BG, U.K.
\and 
Department of Physics, Catholic University of America, Washington DC 20064, U.S.A.
}

\date{Received / Accepted}

\abstract
{The variable X-ray spectra of AGN systematically show steep power-law high states and
hard-spectrum low states. The hard low state has previously been found to be a 
component with only weak variability.  The origin of this component and the
relative importance of effects such as absorption
and relativistic blurring are currently not clear.}
{In a follow-up of previous principal components analysis, we aim to 
determine the relative importance of scattering and absorption effects on 
the time-varying X-ray spectrum of the narrow-line Seyfert 1 galaxy Mrk~766.}  
{Time-resolved spectroscopy, slicing  XMM and Suzaku data down to 25\,ks elements, 
is used to investigate whether absorption or scattering components dominate the 
spectral variations in Mrk 766.}
{Time-resolved spectroscopy confirms that spectral variability in Mrk~766 can be explained by 
either of two interpretations of principal components analysis. Detailed investigation 
confirm rapid changes in the relative strengths of scattered and direct emission or 
rapid changes in absorber covering fraction provide good explanations of most of 
the spectral variability. However, a strong correlation between the 6.97 keV absorption line and 
the primary continuum together with  rapid opacity changes show that 
variations in a complex and multi-layered absorber, most likely a disk wind, 
are the dominant source of spectral variability in Mrk 766.} 
{}

\keywords{accretion; galaxies: active}

\maketitle

\section{Introduction}

Determination of the origin of X-ray spectra of active galactic nuclei has been mired 
in ambiguity, with different classes of candidate models predicting similar 
time-averaged spectra. A  
continuum in the form of a power-law is at the core of all popular models; 
a `reflection spectrum' produced by scattering  from the surface of Compton-thick  
material is also likely to be present, the putative accretion disk
\citep{guilbertrees88,lightmanwhite88,fgf} 
must exist near the continuum region  and such reflection would produce a hard  
spectral component with strong Fe K emission, 
observable below 10 keV  \citep{fgf,zdziarski95,perola02,tanaka95,nandraea97}. 
 Layers of Compton-thin gas 
are also thought to shroud the nuclear system, with strong and (relatively) 
unambiguous evidence 
for the presence of zones covering a large range of 
ionization-states and 
column densities \citep{crenshaw03}. If `competing' models have the same constituents then 
what makes them different? The principal distinction is the degree to which 
reflection processes dominate versus absorption. Also of general interest is 
whether we can detect the signature of 
relativistic blurring in the reprocessed components. 
In practice, models dominated by blurred reflection are indistinguishable from those 
composed of layers of complex absorption. It has proved difficult to progress 
with current data, where one typically has a short baseline for the observation and is limited to 
 model-fitting time-averaged X-ray spectra with only 
modest energy resolution \citep{reevesea05,turnerea05}. 

A possible way forward is to use 
the observed spectral changes over time to break the ambiguity. 
Rather than simply fitting models to time-averaged data we can attempt to disentangle
components that arise on different physical scales and in different zones by their
differing patterns of spectral variability.  Marked spectral variability 
is indeed observed in local AGN down to timescales of thousands of seconds, and the long 
observations of these sources that have been performed recently yield good sampling of 
spectral variability and the possibility of isolating the dominant effects.

The longer observations of AGN, that tend to only be performed once a mission is 
well-established, have turned out to be invaluable because exposure times $>$ a day have 
pushed us into a regime where new results have been observable. 
For example, a long  {\it XMM} observation of Mrk~766 has shown
 that the primary variable component of the spectrum has 
the form of an absorbed power-law, with accompanying ionized reflection
that varies with the continuum.  The latter result  
 first became evident with the discovery of  the correlation 
between ionised line emission and continuum variations down to 10 ks
\citep{miller06}. These results also concur with the 
analysis of the 2001 data by \citet{turner766}, where  evidence for
a periodic Doppler shift of ionised emission from a radius $r \sim 100$r$_g$
was found. 

These long datasets also 
inspired the first applications of principal components analysis (PCA) to data from a single 
X-ray observation \citep{vaughanfabian04}.
By correlating 
flux  at different spectral energies one can decompose data into  
variable constituent components, this method works well to break data into mathematically 
`additive' components (emission components, direct or reprocessed) 
but has a limitation because it cannot separate out the mathematically `multiplicative'
 changes that 
would be caused, for example,  by varying absorption.
The conclusion from the \citet{vaughanfabian04} PCA study of MCG--6-30-15 
was that the spectral variability in that case was driven primarily by 
a power-law of varying normalization combined with a slowly-varying or constant 
hard component of reflection. 

In the first paper in this series \citep{miller07} we investigated spectral 
variability in {\it XMM} observations of Mrk~766, 
a narrow-line Seyfert\,1 galaxy at redshift $z=0.0129$ \citep{osterbrock}.  
We applied a PCA method based on singular value decomposition that 
preserved the full instrumental spectral resolution.
It was found that the source can be modeled adequately using
just two additive spectral components: a 
variable component comprising a power-law and ionised, modestly-broadened Fe emission line 
together with relatively non-variable component(s) that shows  a hard spectral form.  

The possible origins for the hard component have been discussed at
length in Paper I. A model where there is some scattered and self-absorbed 
fraction of the primary continuum  and possibly a
contribution from a distant neutral reflector can
explain the offset component shape and the spectral variability (hereafter the 'scattering' model). 
Alternatively, the hard flux could be a heavily-absorbed fraction 
of the continuum, and we denote this the `partial-covering absorption' model (`p-cov absorption'). 
In the latter model the PCA breakdown into additive components 
indicated that covering fraction changes produce  the observed spectral variability,  
as the additive components do not change much in shape but rather in relative dominance 
as the source flux varies. Importantly, both these interpretations of the PCA work 
without recourse to relativistic blurring of any spectral component. 
 \citet{miller07} discussed fits that include blurred reflection and concluded that significant 
blurring was not required to explain Mrk 766, although 
the source is consistent with the presence of such effects.

Here, we present here a detailed analysis of
time-resolved spectra of Mrk~766, fitting spectral models to individual time intervals.
{\it XMM} has accumulated $\sim 700$ ks of data from this 
highly-variable Seyfert galaxy  from 2000-2005.  The {\it XMM} data
discussed here have already been used to detect variability of the peak energy and 
flux  of a somewhat broad, ionised Fe emission line in the $6-7$\,keV range 
\citep{miller06}. Here we show, for the first time, the  PCA model fit to the 
data themselves, comparing the two physical interpretations of the 
PCA results presented by \citet{miller07}, i.e. the scattering and p-cov absorption models, 
to understand whether scattering/reflection or absorption effects are dominant.
 Of particular interest here 
is a follow-up of the time-dependent behavior of two absorption features found in the Fe K-band  using PCA,  
a feature at 6.97 keV likely originates in H-like Fe while a feature at 7.3 keV is of uncertain origin.  

In this paper, we concentrate 
on the origin of spectral variability in Mrk 766. In this sense, consideration of what we 
dub the `scattering model' may be taken as including consideration of a 
relativistically blurred reflection model, since the scattered component may also be
fitted by such a model \citep{miller07}, although we note in section\,\ref{discussion} how
absorption line variations 
may be used to distinguish relativistically-blurred from
non-blurred models.
Whatever the model fit to the `hard' component, its temporal behavior is a key 
piece of information that may help us disentangle the possible explanations for
its origin.
In addition to this new treatment of the {\it XMM} data we present new {\it Suzaku} 
data allowing us to extend the analysis 
to $\sim 50$ keV, and increase the temporal baseline for the study.

\section{The Data}

\subsection{XMM-Newton data}

In this paper we utilise all {\it XMM-Newton} (hereafter {\it XMM}) European Photon Imaging 
Camera  (EPIC) pn CCD \citep{struder} data available 
for Mrk~766, covering  2000 May 20 (science observation ID 0096020101),
2001 May 20-21 (science observation ID 0109141301)
and 2005 May 23 (science observation IDs in the range 0304030[1-7]01).
The data reduction is described by \citet{miller06} and as in that work, 
the combination of some non-optimal
modes for the Metal Oxide Semi-Conductor CCDs, photon pileup and inferior signal-to-noise led us to use
only the pn CCD data in this analysis.

To perform the time-resolved spectroscopy we found that even
time-selections of 25\,ks duration gave adequate signal-to-noise
in all spectra (this yielded slightly lower effective exposure times
in some cases, due to data gaps). This sampling gave a good match to
the variability timescales evident in the data.  Where datasets did
not divide exactly into 25 ks bins, the final bin of an `obsid'
(orbital segment of data) was allowed to be slightly shorter or longer
than this.  This yielded 28 spectral slices across all available data;
2000 May yielded two spectral slices, 2001 yielded five slices and
2005 yielded 21 spectral slices.  With time-sliced spectra, despite
our attempts to optimize the selections, the signal-to-noise at high
energies, $E > 7.5$\,keV, was still found to be low.  To optimize the
spectral fitting we binned all spectra consistently with energy bins
equal in width to the half width half maximum (HWHM) spectral resolution 
of the EPIC pn instrument.  The HWHM varies
with photon energy, so the bin widths chosen also vary continuously with energy,
from $\sim 35$\,eV at $E \sim 1$\,keV to $\sim 80$\,eV at $E \sim
10$\,keV.

The energy range chosen for the analysis initially is $1.0-9.8$\,keV,
this focuses the study on gas closest to the nucleus and avoids
modeling in detail emission/absorption lines below 1 keV, some of
which likely arise from gas relatively distant from the central
engine (e.g. \citealt{og03}).  The truncation at 9.8\,keV is 
where signal-to-noise reaches a very low level.

\subsection{Suzaku data}

Mkn 766 was observed by {\it Suzaku} from 2006 November 16 00:34:54 to 2006  
November 18 08:44:01, in XIS nominal mode.  We used events files from  version 1.3.2.6 
of the {\it Suzaku} pipeline.  

The XIS data was reduced using tasks from v.6.1.2 of the `ftools' package. It
was screened with XSELECT to exclude data within the  South Atlantic Anomaly (SAA) and 
data within 436 seconds of the SAA  (using the T\_SAA parameter).  Additionally we 
excluded data with an Earth elevation angle less than 5$^\circ$ and Earth day-time 
elevation angles less than 20$^\circ$.  Finally we selected good  events with grades 
0,2,3,4, and 6 and removed hot and flickering  pixels using the SISCLEAN script.  This
 screening left us with  effective exposures of $\sim 87$ ks for each of the 
XIS 0,  1, and 3 chips.  Note that the spaced-row charge  injection (SCI) 
was not active during this observation. The XIS products were extracted from 
circular regions of 2.9\arcmin  while background spectra were extracted from a region 
of the same size offset from the source (and avoiding the chip corners
with the calibration sources).  The response and ancillary response
files were then created using the tasks XISRMFGEN and XISSIMARFGEN,
respectively.  The source contributed 98.6\%, 95.3\%, and 98.4\% of
the total counts in from the XIS 0, 1, and 3 products, respectively.
The orbit of {\it Suzaku} leads to the exposure noted above being
spread out over a long, 200\,ks, baseline.  To obtain time slices with
similar exposure to the {\it XMM} spectra we split the data into four
even time periods covering a baseline of 50\,ks each, for spectral
analysis.

The PIN background events file was provided by the HXD instrument
team, used in conjunction with the source events file to create a good
time interval applicable to both the source and background.  The
background events file was generated using ten times the actual
background count rate, so we increased the effective exposure time of
the background spectra by a factor of 10.  We found the
deadtime correction factor using the HXDDTCOR task with the extracted
source spectra and the unfiltered source events files.  After deadtime
correction, we were left with an effective exposure of $\sim 87$\,ks.
To take into account the cosmic X-ray background 
\citep{boldt87,gruber99}, Xspec version 11.3.2 \cite{arnaud}
was used to generate a spectrum
from a CXB model, normalized to the {\it Suzaku} field of view, and combined
with the PIN simulated background file using MATHPHA to create an
total background file.  The source contributed 6.7\% of the total
counts.

\section{The Models}

In this
paper, as we have many spectral components in the fits, we refer to
the ionised reflection that is viewed directly (and appears from the
variability of the Fe\,K emission line to correlate with
the directly-viewed powerlaw) as the `ionised reflector'. We denote
the hard/offset component, 
as the `hard' or `scattered' component. It is not clear whether the scattering gas
is Compton-thick in that case and the use of the term `scattered' allows an
easy distinction between this steady component and the ionised
reflector that is correlated with the continuum flux.
\citet{miller07} also considered whether models where the spectral
variability was dominated by absorption changes might explain the
data, concluding that the hard component could be described as an
absorbed fraction of the primary continuum and that changes in
covering fraction could then explain the source behavior.  
As described in the introduction we term this
the `p-cov absorption' model.
Both `scattering' and `p-cov absorption' models are consistent with the PCA
provided the spectral variability is
dominated by covering fraction changes in the p-cov absorption model. 

The general approach taken was to utilize the fits to the direct and
offset components from the application of PCA to the {\it XMM} data
\citep{miller07}, then extend the analysis to include the broader bandpass {\it Suzaku}
data, as a test of the model on data not used to create it.  

\citet{miller07} fitted the spectral elements derived from PCA, to investigate the
source in terms of additive spectral components.  Here we fit the data
directly using time-resolved spectroscopy  and this allows us to test 
how well the two PCA interpretations  fit each 
time period for Mrk~766, and to see how the different model parameters vary with
time. This analysis 
also aims to expand upon the PCA by finding an explanation for the
residual spectral variability that was not well modeled
there.  As noted above then, we apply two interpretations of the PCA decomposition: first we
fit the `scattering model' and then the `p-cov absorption
model' that uses varying covering fraction to explain the source. The
derivation of these models from the PCA is discussed in detail by
\citet{miller07}, and so the two models to be utilized are simply
summarized below. Here we use the $\xi$ form of ionization parameter where  
$\xi = L/nr^2$, $L$ is the 1-1000 Rydberg luminosity, 
$n$ the gas density and $r$ the absorber-source distance. 
We quote $\xi$ in units of erg\,cm\,s$^{-1}$ throughout. 

\subsection{The Scattering Model}

The scattering model  (Fig.\,\ref{pic}) is that found from fitting the PCA hard component using 
the scattered primary continuum (whose slope was determined from fitting the PCA variable component). 
The full spectral model then consists of: 

\begin{enumerate}

\item  A power-law of constant slope $\Gamma=2.38$ and variable normalization. 
\item  An ionised reflector  
 that varies  with normalization tied to that of the power-law. 
\item  A $\sim$constant scattered component. This is represented by 
summing components 1 and 2 and then allowing a scalar to 
represent the scattered fraction. 

\item   A complex layer of  ionised absorption in front of 
 the scattered component, represented by the cmbination of two layers:  
 $N_H(1)=4.74 \times 10^{22} {\rm cm^{-2}}, \log\xi(1)=-0.17$, 
 $N_H(2) = 3.97 \times 10^{23} {\rm cm^{-2}}, \log\xi(1)=3.05$.  
This physical might arise if the reflecting/scattering surface 
has its own atmosphere, for example.
\item  A neutral  reflector with no blurring 
using  the {\sc xspec}  pexrav model with  a Gaussian 
emission line representing Fe K$\alpha$ at the 
strength expected for a slab of material inclined 
at 30$^{\rm o}$ \citep{fgf}.
\item  A low-column, low-ionization absorber fully covering all components: 
 $N_H(3) = 2.77 \times
10^{21} {\rm cm^{-2}}, {\rm log} \xi(3)=1.80$ covered all components
during 2000-2001 while $N_H(3) = 2.41 \times 10^{21} {\rm cm^{-2}},
{\rm log} \xi(3)=1.26$ covered all components during 2005 and during
the 2006 {\it Suzaku} observation.

\item  Two Gaussian absorption lines, allowed to float in normalization, with 
energies fixed at 6.97\,keV and 7.3\,keV (in the rest-frame of the host galaxy).

\end{enumerate}

Construction of this model assumed the highest state observed to date
(in 2001)
to represent the sum of directly viewed and reflected components; data from the
lowest flux state observed in {\it XMM} data dominated the
determination of the form of the hard component, although a small
contribution from the direct component was allowed in that
determination: the initial parameters values were chosen from the 
model fits to the PCA components.

\subsection{The P-Cov Absorption Model}

\begin{figure}

\begin{minipage}{88mm}{
\resizebox{88mm}{45mm}{
\rotatebox{0}{
\includegraphics{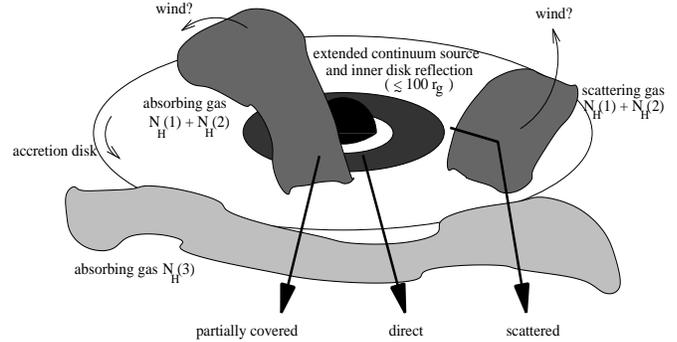}
}}}
\end{minipage}
\caption{A schematic representation of the inner regions of the models 
described in Section 3. 
The  distant 
 reflector is  omitted from the scattering model, for clarity}  
\label{pic}
\end{figure}

The p-cov absorption model (Fig.\,\ref{pic}) is determined using several layers of ionized gas
to explain the overall shape of the hard component, this complex
absorption profile for the absorbed fraction is consistent with 
many physical models. The model is composed of:

\begin{enumerate}
\item A power-law of constant slope $\Gamma=2.18$ and variable normalization
\item  An ionized reflector 
 that varies with normalization tied to that of the power-law 
\item 
A large and complex layer of absorption composed of   
 $N_H(1) = 4.03 \times 10^{22} {\rm cm^{-2}}, \log\xi(1)=1.19$
overlaying a second layer 
$N_H(2) = 4.57 \times 10^{23} {\rm cm^{-2}}, \log\xi(1)=2.3$ comprised the complex absorber whose 
 variable covering fraction accounts for much of the observed spectral variability. 
The column densities and ionisation 
parameters of these two layers were initially assumed to be constant
throughout the time-resolved fitting for both models but later we
found it necessary to allow $N_H(1)$ to vary (Table 1). 

\item A narrow, neutral component of Fe K$\alpha$   
\item A low-column, low-ionisation absorber covers all components: 
 $N_H(3) = 5.38 \times 10^{21} {\rm cm^{-2}}, \log
\xi(3)=2.27$ for 2000-2001 $N_H(3) = 3.91 \times 10^{21} {\rm
cm^{-2}}, \log\xi(3)=1.93$ for 2005-2006.

\item Two Gaussian absorption lines, allowed to float in normalization,
 with energies fixed at 6.97 keV and 
7.3 keV (in the rest-frame of the host galaxy).
\end{enumerate}

Construction of this p-cov absorption model assumed the highest state observed to date
to be uncovered with respect to this complex dense absorbing layer,
although the low column, moderately ionized gas layer of component 5
covers the direct component (and all other components) at all times.
The power-law index is flatter in this model than in the scattering
model because in the latter the hard scattered component exists in
all flux states, but it is absent in the p-cov  absorption model. 
The high-state data thus determined the form of the underlying
continuum before the variable covering by absorption is applied. The
fits then allow any additive combinations of the uncovered and covered
fractions to fit each spectral slice. The fact that almost all shape
parameters are fixed means the model has very few degrees of freedom
and the method of construction is designed to isolate the key
variables most effectively.  The uncovered and covered fractions are
naturally anti-correlated with this construction, although they
have not been strictly required to always sum to the same total
intrinsic flux
(i.e. some intrinsic variation in continuum normalization is allowed
but not required).

\subsection{Fitting the models}
Thus, following the results from the PCA, the slope of the power-law in each
model was frozen at the best fit value found  in each scenario. The
ratio of power-law to ionized reflector was also frozen for each model
and the normalization of the summed pair 1 plus 2 allowed to vary;
this reproduces the known line/continuum correlation \citep{miller06}. 
It was found that the low column, low-ionization 
layer of absorption  needed to have a different column density
for the 2000-2001 
and the 2005 epochs, and the column density was fixed at the best-fit 
value for each of those two time periods, in each model run. 

The 2005 value derived from 
{\it XMM} was later found applicable to the {\it Suzaku} data and
utilized in those fits.

\begin{figure}

\begin{minipage}{88mm}{
\resizebox{88mm}{45mm}{
\rotatebox{0}{
\includegraphics{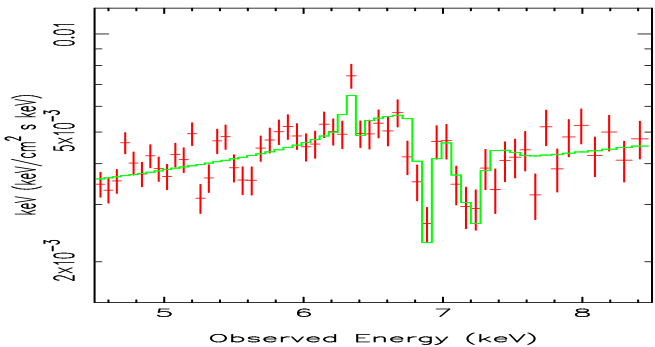}
}}}
\end{minipage}

\caption{A p-cov absorption fit to a slice of the low-state XMM data (slice 9) 
illustrating the presence of the two absorption features, at 6.97 keV and 
7.3 keV in the rest-frame of the host galaxy}  
\label{fe_zoom}
\end{figure}

In both cases, the ionized reflector was modeled using the 
`reflion' model of \citet{rossfabian}, assuming solar
 abundances and that $\xi$ was fixed at 1600\,erg\,cm\,s$^{-1}$, as determined 
from fitting the mean spectrum. Some  blurring was allowed 
in the reflection spectrum via 
a \citet{laor91} convolution model, provided in xspec by the `kdblur' 
function, this was a very modest blurring constrained to arise
 within 100-300\,$r_g$ from a disk inclined at 
30$^{\rm o}$ to the line-of-sight, and necessary to model the 
$\sigma \sim 0.27$ keV width of 
 the ionized diskline. 

Layers of absorption were modeled using tables generated from 
{\sc xstar} version 21ln  \citep{kallman04} 
with solar abundances,  and turbulent velocity 200\,km\,s$^{-1}$. 
The Gaussian model absorption lines at 6.97 and 7.3\,keV were 
included in addition to the complex absorption components  as 
these were not modeled by those absorption layers. The 6.97\,keV line 
most likely arises from highly ionized 
gas and thus would have little or no signature at lower energies 
(unless the column density is enormous). 
The 7.3\,keV line is of 
uncertain identification and thus it is most safely modeled in an 
isolated manner. 

For the scattering model, the initially free components were:
 the normalizations of the directly-viewed and scattered 
 fractions of the summed powerlaw plus ionized reflector (i.e. 1 plus 2). 
Additional free components were:
the normalization of an additional neutral component of reflection, modeled using 
the {\sc xspec} pexrav model with a linked Fe\,K$\alpha$ line of strength expected 
from a neutral reflector 
subtending 2$\pi$ steradians to the illuminating source and viewed at 30$^{\rm o}$  
\citep{fgf} and the normalizations 
 of narrow Gaussian components to model absorption features 
at  6.97 and 7.3 keV (in the rest-frame of the host galaxy). 

For the p-cov absorption model the initial  free components were: 
 the normalizations of the unabsorbed and absorbed fractions of the spectrum  
(by design, these are anti-correlated in this model); and the Gaussian line 
normalizations, as before. In this model the neutral Fe line at 6.4\,keV is not linked to a 
component of neutral reflection, but allowed to vary in isolation; this model assumes that 
line arises from transmission through the absorber and not from a neutral reflector. 

The absorption lines, while highly significant in the
mean {\it XMM} spectrum and in the PCA offset component
\citep{miller07} are highly sensitive to the form of the spectral
model, as they fall in a region of some complexity. For this reason,
the line fluxes derived from
each model differ slightly, 
and in the paper we present both sets of values for
comparison. A close-up of the Fe\,K region is shown in
Fig.\,\ref{fe_zoom} for slice 9 of the {\it XMM} data.

\begin{figure}
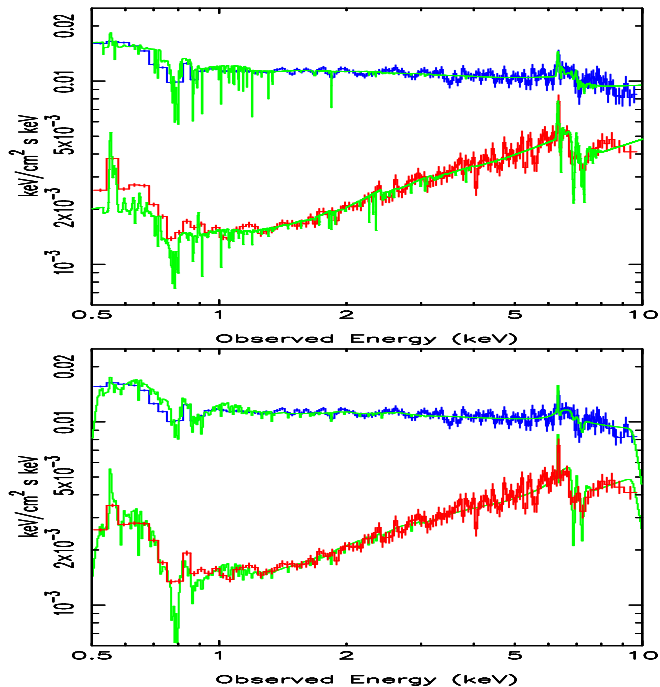


\begin{minipage}{88mm}{
\resizebox{88mm}{45mm}{
\rotatebox{0}{
\includegraphics{new2a_scatt.ps}
}}}
\end{minipage}
\begin{minipage}{88mm}{
\resizebox{88mm}{45mm}{
\rotatebox{0}{
\includegraphics{new2b_pc.ps}
}}}
\end{minipage}

\caption{Scattering model (top) and p-cov 
absorption  model (bottom) fits shown as 
green solid lines 
representing fits to  two 
of the individual time slices of
{\em XMM-Newton} data (slice 6, blue; slice 9, red). 
The agreement between data and model  illustrates the 
applicability of either model 
over the broad range of flux covered by the source during the {\it XMM} 
observations. The model lines are shown extrapolated down to 0.5 keV illustrating 
how exclusion of the softest data from the fits has not compromised 
our ability to distinguish 
models; the soft data indicate a preference for the partial-covering 
absorption model}  
\label{xmm_scatt_pc_xmmfits}
\end{figure}

\section{Results} 

\subsection{XMM results} 
 
Fitting first all 28 time slices of
the {\it XMM} data over 1-9.8\,keV in the observed frame, the scattering model
gave a total goodness-of-fit $\chi^2=4796$ for 3752 
degrees of freedom (d.o.f.) while the
p-cov absorption model gave 3785 for 3752 d.o.f..  
The fits confirm that despite having relatively few varying components,
both our two physical  representations of the PCA components
do indeed describe the spectral variability to first order, and the models 
describe many spectral slices 
completely satisfactorily (Fig.\,\ref{xmm_scatt_pc_xmmfits}).  
Successful first-order application of the two physical models to the
 data leaves us in a position to now determine the origin of the
 additional complexity that could not be accounted for from the PCA. This
 is a crucial step because we need to achieve a statistically superior
 set of fits before a meaningful comparison of models can be made.  To
 this end, we examined the residuals to the worst fits, finding that
 allowing a change in column density of one of the zones of absorption
 can explain the remaining residuals.  Allowing this extra degree of
 freedom significantly improves fits to both models.  This makes sense, since 
 the PCA could not handle multiplicative changes to the source
 spectrum (note that covering fraction changes alone would show up as
 additive changes in PCA, since the profile doesn't change but rather the amount of covered and uncovered 
source spectrum) and one might thus expect that a
 multiplicative variation would be the origin of spectral variability
 not handled by PCA. The column density changes are a natural
 modification as both models incorporate significant columns of gas.
 Allowing the ionisation parameter of that zone to be free in addition
 to the column did not further improve the fits, so we did not allow
 $\xi$ as a free parameter.  The degeneracy between $N_H$ and $\xi$
 means that while we have modeled these residual variations as column
density variations as tabulated, it is impossible
 to determine how much of this change is actually due to column changes and how
 much is due to ionisation changes.

An absorber in the line-of-sight would respond to changes in the
continuum flux with recombination timescales down to seconds for gas
in the density regime likely for layers close to the nucleus, and even
faster ionisation timescales. In the scattering model our absorption
happens after scattering, so there may be significant time lags and
we would not necessarily expect to see $\xi$ changes consistent
with the continuum variation. In the p-cov absorption model the gas might be
expected to respond to the continuum variability but it is not
possible to unambiguously determine how much of the continuum
variability is due to the covering variations and how
much to intrinsic continuum flux changes.  Given these ambiguities 
we do not seek consistency between observed flux changes and the
ionisation state of the absorbers.

After freeing the column density for $N_H(1)$ the fits were  re-run 
with the column density of $N_H(1)$ allowed to float in addition to
the free parameters already noted.  Freeing this column improved both
models: the scattering model was improved to an overall 
goodness-of-fit $\chi^2=4274$ for 3724 d.o.f. while the p-cov 
absorption model yielded 3488 for 3724 d.o.f. for the {\it XMM} data.  
The improvement to the fits in both
cases indicates that column density changes are a significant component 
of spectral variability in Mrk 766 in either the scattering or absorption scenarios. 
While the absorption model provides a fit that appears statistically
superior in the current model constructions, we note that the
scattering model has been constructed using basic model components to
attempt to describe an interesting physical scenario. The recourse to
this approximation for the combined scattering and self-absorption
effects that may occur in (for example) a disk wind, is due to the
lack of a full self-consistent wind model at this time. For this
reason, the simple fit statistic should not be a basis for ruling out
this crudely-parameterized but interesting physical possibility.

\begin{figure}
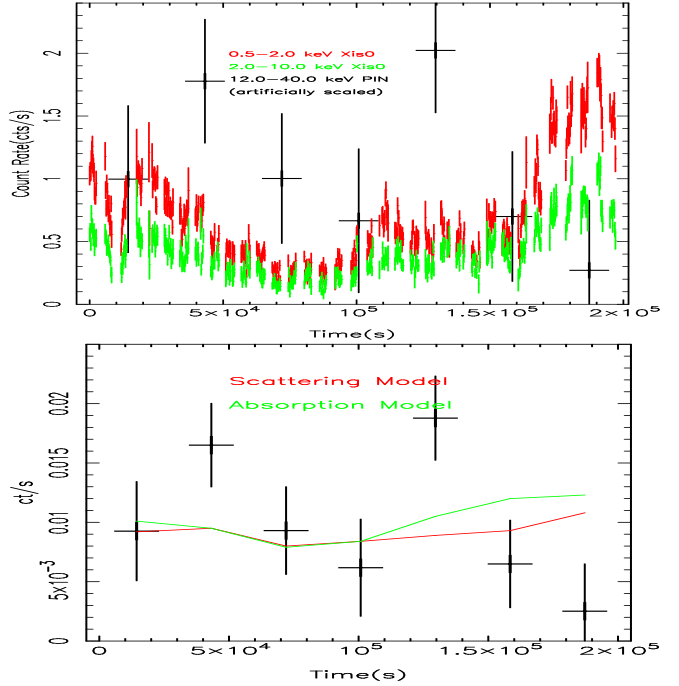

\begin{minipage}{88mm}{
\resizebox{88mm}{45mm}{
\rotatebox{0}{
\includegraphics{xis_pin_lc.ps}
 }}}
 \end{minipage}
 \begin{minipage}{88mm}{
 \resizebox{88mm}{45mm}{
 \rotatebox{0}{
 \includegraphics{pin_pred.ps}
}}}
\end{minipage}
\caption{
Top: XIS0 time series in 500 s bins in the 
0.5-2.0 (red)  and 2-10 keV (green)  bands 
and from the PIN data covering $\sim 12-40$ keV (black)
Bottom: PIN light curve versus predicted flux in the 12-40 keV band  
from the scattering (red) and absorption (green) models. 
}

\label{suzaku_lc}
\end{figure}

\subsection{Suzaku results}

Extending our consideration to the {\it Suzaku} data we first created
a light curve from the PIN data, and compared it to light curves
in the 0.5-2 and 2-10 keV bands of XIS0.  Fig.\,\ref{suzaku_lc} shows
those time series and it appears that while the XIS
data vary by a factor $\sim 6$ from trough to peak, the PIN data are inconsistent with a zero-lag correlation in 
variability. However, there are some uncertainties in the production of the 
PIN light curve that relate to the difficulty in producing an accurate subtraction of the 
instrumental background level: for this reason the PIN curve is not fitted to the model predicted 
rates, although these are illustrated in Fig.\,\ref{suzaku_lc}. 
%We also compared the PIN curve to a 
%constant model and to fluxes 
%in the PIN bandpass from the two models. A fit to a constant model gave 
%$\chi^2_r=2.46$ for 6 d.o.f., a fit to 
%the scattering model predicted 
%curve yielded reduced $\chi^2_r=2.94$ for 6 d.o.f. and a fit to the 
%p-cov absorption model prediction gave $\chi^2_r=3.28$ for 6 d.o.f. The 
%scattering and absorption 
%models are inconsistent with the  PIN variability at significance level $\la 1$\%, 
%but neither is significantly worse than the other. 

\begin{figure}
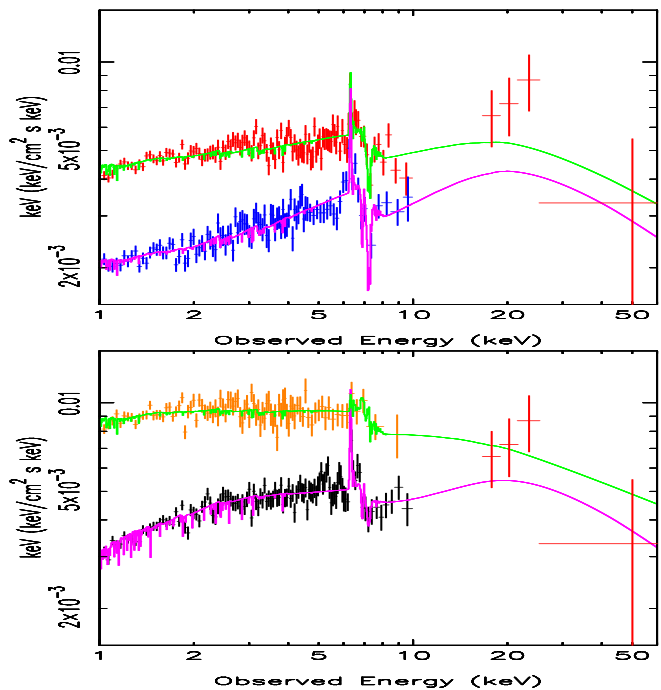


\begin{minipage}{88mm}{
\resizebox{88mm}{45mm}{
\rotatebox{0}{
\includegraphics{sc_panel1.ps}
}}}
\end{minipage}
\begin{minipage}{88mm}{
\resizebox{88mm}{45mm}{
\rotatebox{0}{
\includegraphics{sc_panel2.ps}
}}}
\end{minipage}

\caption{Scattering model fits to the individual time slices of
{\em Suzaku} data.Top:  Red is slice 1, blue is slice 2; Bottom: 
Black is slice 3, Orange is slice 4. PIN data have been adjusted downwards 
by the 13\% flux correction determined from simultaneous 
fitting of XIS and PIN spectra}
\label{suz_sc_fits}
\end{figure}

\begin{figure}
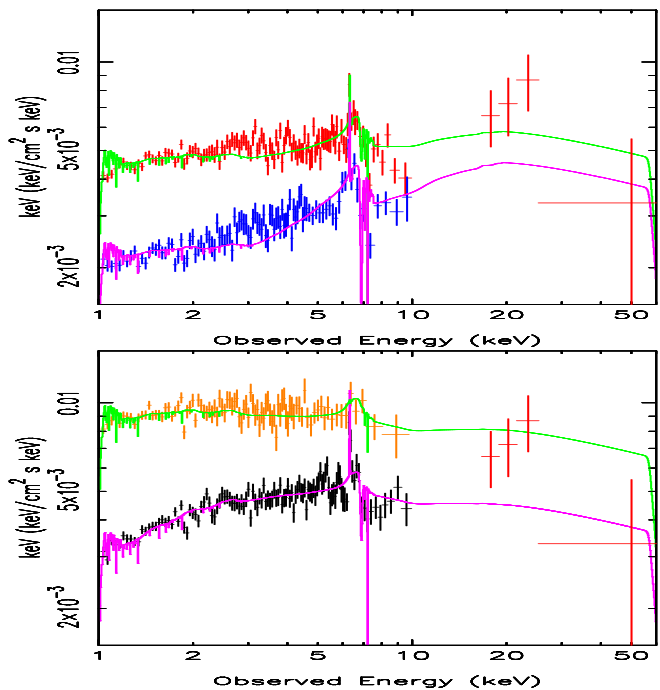


\begin{minipage}{88mm}{
\resizebox{88mm}{45mm}{
\rotatebox{0}{
\includegraphics{pc_panel1.ps}
}}}
\end{minipage}
\begin{minipage}{88mm}{
\resizebox{88mm}{45mm}{
\rotatebox{0}{
\includegraphics{pc_panel2.ps}
}}}
\end{minipage}

\caption{P-cov absorption model fits to some of the individual time slices of
{\em Suzaku} data, color coding and PIN data as for Fig.\,\ref{suz_sc_fits}
}
\label{suz_pc_fits}
\end{figure}

When the PIN data are binned into the four time intervals chosen for spectral analysis 
 the light curve is consistent with a constant flux and so  we  used the mean PIN spectrum to obtain a fit 
with each XIS time slice (Tables A1,2; Figs.\,\ref{suz_sc_fits},\,\ref{suz_pc_fits}).

Fitting the spectral data  we found that three of four time slices 
were well explained in the same way as the {\it XMM} data (Tables A1,A2). 
However, the third time slice was 
a very poor fit to both models (reduced $\chi^2_r=1.88$ for the scattering model, 
$\chi^2_r=2.99$ for the 
p-cov absorption model) until the second large gas column, $N_H(2)$, was also allowed to 
float. Allowing a second layer of absorption 
to vary yielded significantly better  fits to both models (Tables A1,A2) although 
yet more complexity is apparently indicated (Fig.\,\ref{suz_sc_fits}, \ref{suz_pc_fits}).

Both physical interpretations of the PCA results provide reasonable
fits to all {\it XMM} spectra. Tables~A1 and A2 show the fit parameters
for each model, along with the goodness of fit for each
slice. Fig.\,\ref{compflux} shows the primary parameters for the fitted models, 
i.e. the flux of the two dominant spectral components. 
Comparison of Tables A1 and A2 shows that the p-cov absorption model
provides the statistically superior fit to most time slices sampled
from {\it XMM} while the scattering model is the best fit to {\it
Suzaku} data. However, the aim here is to determine whether
scattering/reflection or absorption effects dominate the spectral
variability and, as noted above, if our template models are not
perfect representations of the physical scenarios then we might be
misled by a simple comparison of the $\chi^2$ statistic. To understand the
nature of the source variations we need to consider which of the sets
of results makes the most physical sense, and part of this involves
consideration of whether any obvious parameter correlations emerge.

\subsection{Line Variability}

\begin{figure}
\begin{minipage}{88mm}{
\resizebox{88mm}{!}{
\rotatebox{0}{
\includegraphics{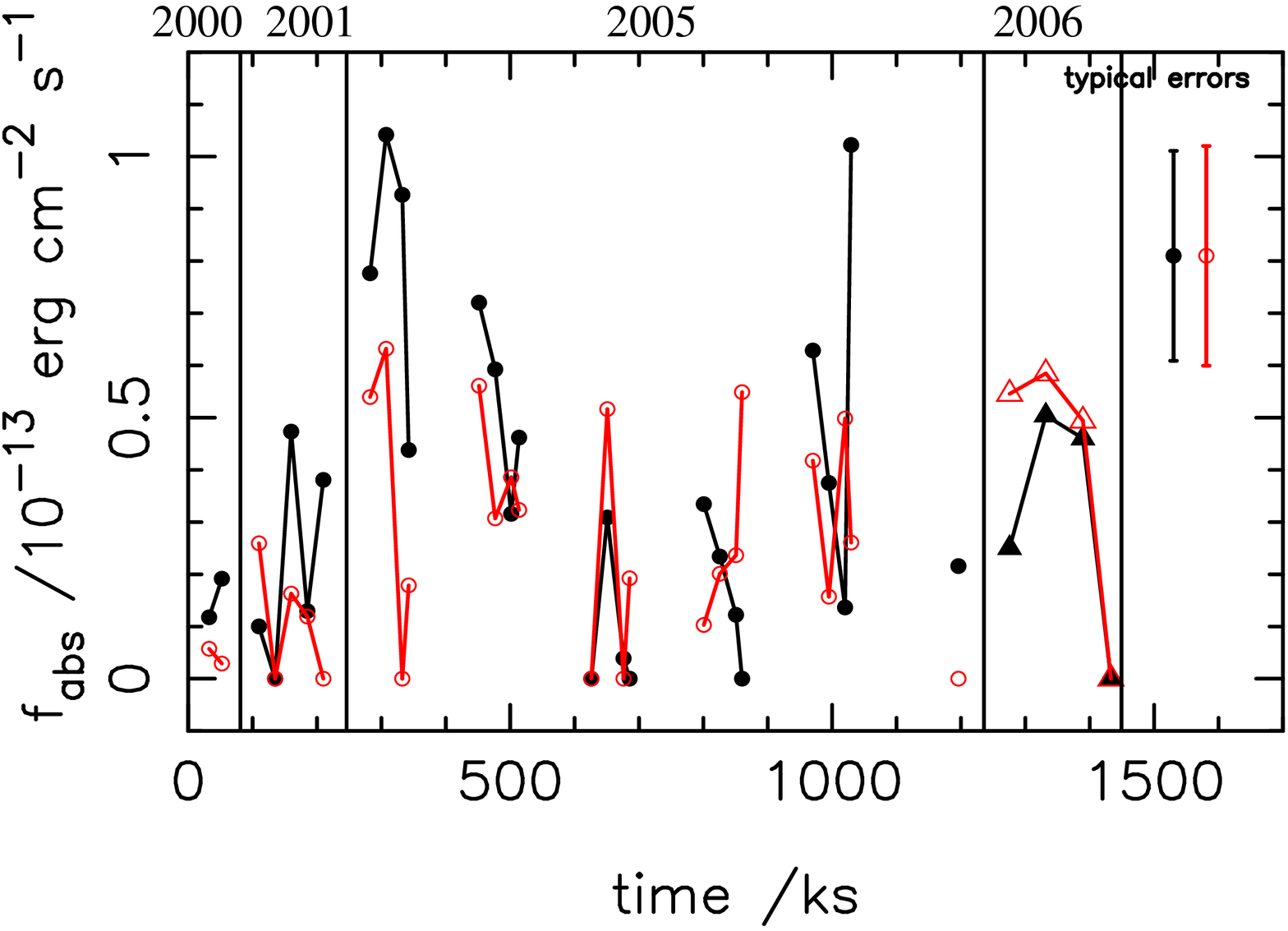}
}}}
\end{minipage}

\begin{minipage}{88mm}{
\resizebox{88mm}{!}{
\rotatebox{0}{
\includegraphics{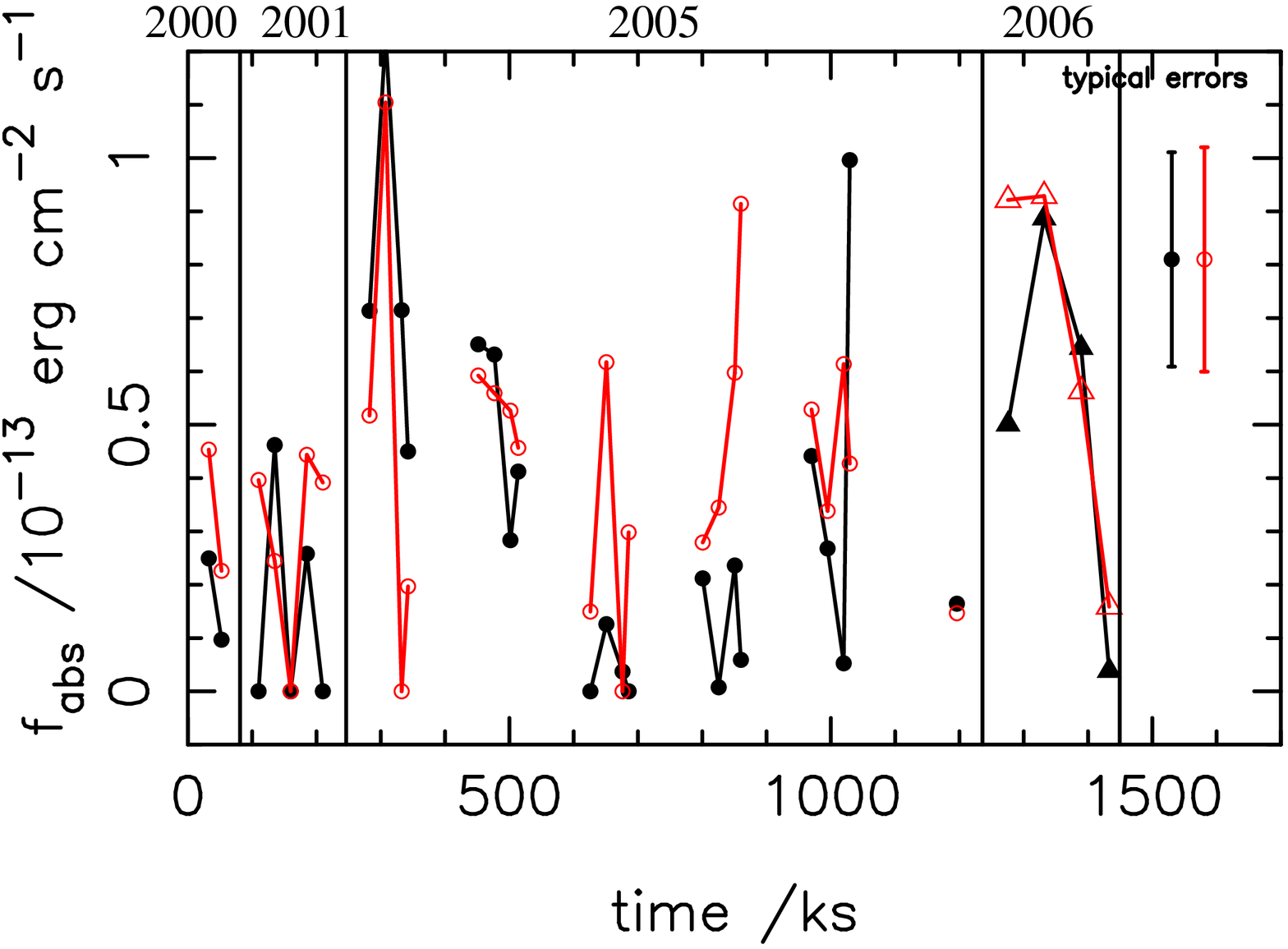}
}}}
\end{minipage}
\caption{Absorption line fluxes for lines at 7.3 keV (red, open symbols) and 
6.97 keV (black, solid symbols) from the scattering (top) and partial-covering (bottom) fits. 
}
\label{linecomp}
\end{figure}

Fig.\,\ref{linecomp} shows the absorption line fluxes plotted against
time.  The 6.97 keV line appears to show significant variability while
the 7.3 keV line shows less indication of any variation; to quantify
this we performed a $\chi^2$ test against a model representing a constant line flux.  The
fitted values of line flux were slightly different for the two model
applications because the line depths are critically dependent on the
placement of the continuum.
 
In the scattering model, the 6.97 keV line flux variations yielded 
$\chi^2=66$/31 d.o.f, variable at significance level $\sim 2\times 10^{-4}$, 
the 7.3 keV line flux variations gave $\chi^2=28$/31 d.o.f., consistent 
with a constant value. 
In the p-cov absorption model, the 6.97 keV line yielded 
$\chi^2=71$/31 d.o.f, also variable at significance level $\sim 5 \times 10^{-5}$, 
but the 7.3 keV line gave $\chi^2=52$/31 d.o.f., variable at 
significance level $0.01$. 
In this case the evidence for variation in the 7.3 keV line 
turns out to be due to one 
deep line measurement during the low state in spectral slice 9, 
exclusion of that single value shows the rest of the line fluxes to be 
consistent with a constant value. 

In spite of the dependence of the lines on the continuum model, the
absorption line at 6.97 keV shows significant variability in flux for both of
the different parameterizations of the data. This strengthens our
confidence in the robustness of the line variability result in this
case.  However the discrepant conclusions on the 7.3 keV absorption
line in the two scenarios leads us to conclude we cannot be sure of
the variability behavior of this line component.  However,  the equivalent width of 
the 7.3 keV line does vary systematically with the total source flux and
is not consistent with being a line of constant equivalent width.

The {\em emission} line at 6.4 keV also has different implied behavior in
the scattering and p-cov absorption models. In the former it is tied to the
neutral reflector and measures of the line flux yield $\chi^2=70$/31
d.o.f (variable at significance level $8\times 10^{-5}$) while in the p-cov 
absorption model
it is not linked to any other continuum component and yields
$\chi^2=8.3$/31 d.o.f, consistent with a constant value. The question
of variability of this line component remains ambiguous, while the Fe
emission varies in this regime it is impossible to disentangle the
neutral line from the wings of the emission line from the ionized
reflector and the origin of the neutral line cannot be conclusively
determined.

\section{Parameter Correlations}

The presence or lack of compelling correlations between 
parameters could help us distinguish 
between the scattering and partial-covering models. 
The achievement of reasonable fits  (Tables A1, A2)  means reliable 
parameters can now be extracted and we can investigate the correlations 
between them for each of the models. 

\begin{figure}
\begin{minipage}{88mm}{
\resizebox{88mm}{!}{
\rotatebox{0}{
\includegraphics{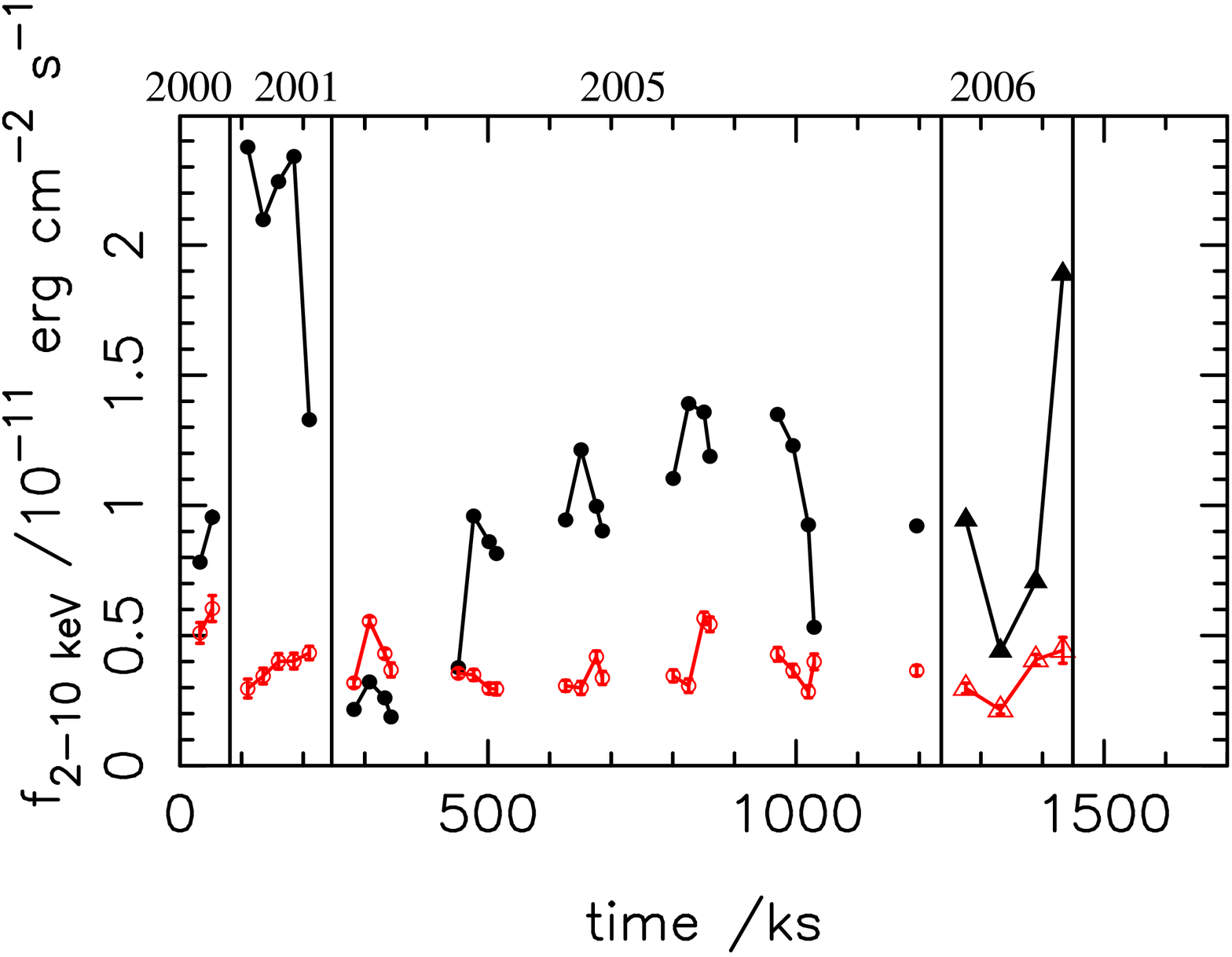}
}}}
\end{minipage}
\begin{minipage}{88mm}{
\resizebox{88mm}{!}{
\rotatebox{0}{
\includegraphics{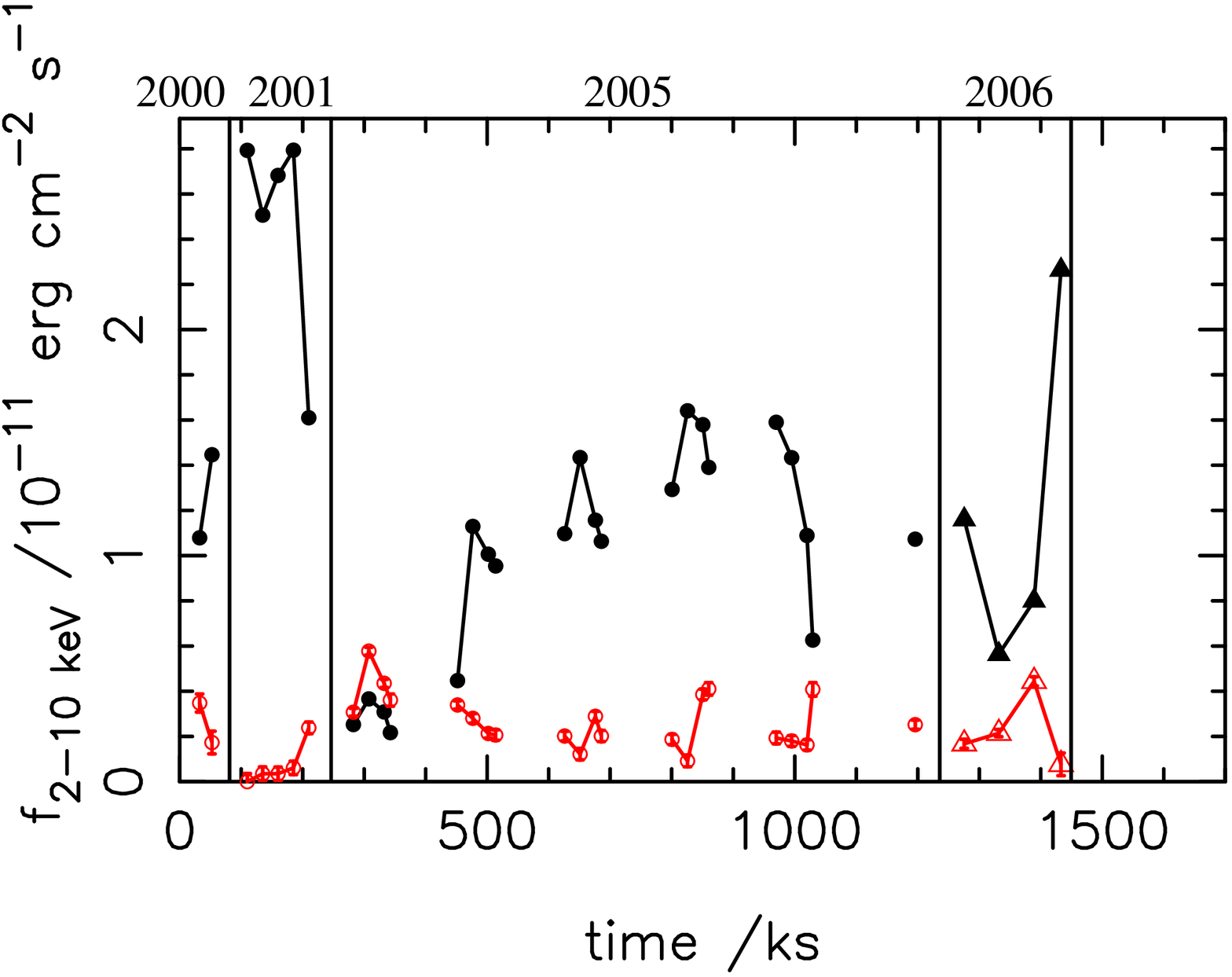}
}}}
\end{minipage}
\caption{
Top: Direct (black, solid symbols) and scattered (red, open symbols) component
 amplitudes of the source flux for 
the scattering model. Bottom: Direct (black, solid symbols) and absorbed (red, open symbols) 
component amplitudes of the source flux from the partial 
covering model. Solid vertical lines 
show where long breaks in time have been omitted for clarity. 
}
\label{compflux}
\end{figure}

\begin{figure}
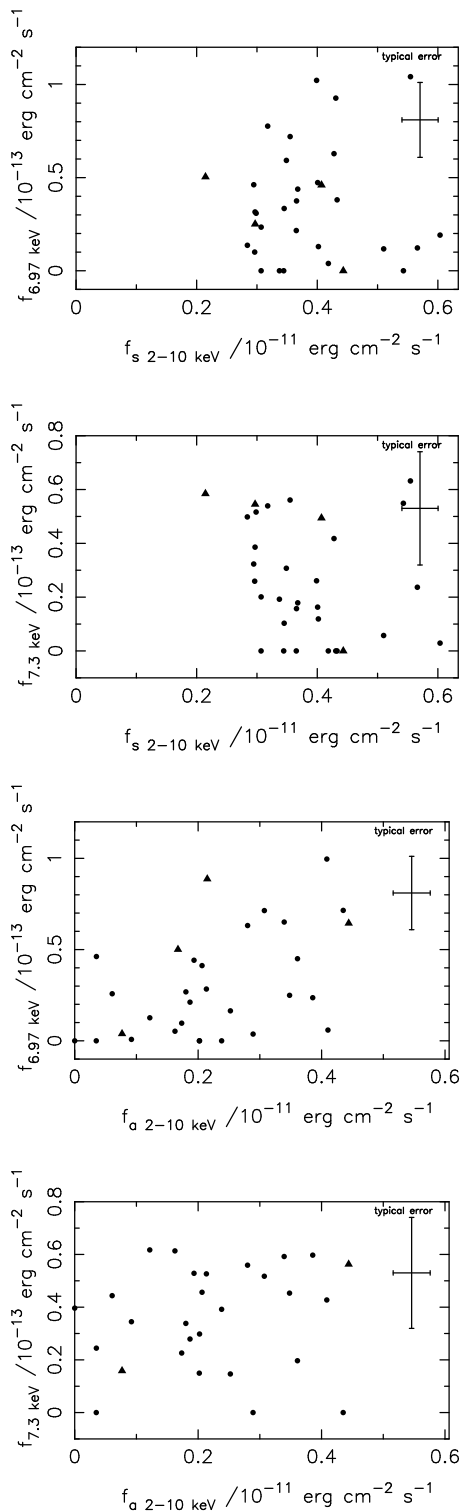

\begin{minipage}{60mm}{
\resizebox{60mm}{!}{
\rotatebox{-90}{
\includegraphics{sc4.ps}
}}}
\end{minipage}

\begin{minipage}{60mm}{
\resizebox{60mm}{!}{
\rotatebox{-90}{
\includegraphics{sc5.ps}
}}}
\end{minipage}

\begin{minipage}{60mm}{
\resizebox{60mm}{!}{
\rotatebox{-90}{
\includegraphics{pc4.ps}
}}}
\end{minipage}

\begin{minipage}{60mm}{
\resizebox{60mm}{!}{
\rotatebox{-90}{
\includegraphics{pc5.ps}
}}}
\end{minipage}
\caption{Absorption line fluxes 
 versus the hard component in the 
scattering model (top two panels) and  
 p-cov  model  (bottom two panels) for {\it XMM} (dots) and 
{\it Suzaku} (triangles) data. }
\label{line-hardcomp-correlations}
\end{figure}

First we plot the most robustly-measured parameters, the directly-viewed 
and scattered or absorbed fluxes. 
Fig.\,\ref{compflux} shows the directly viewed flux (powerlaw plus correlated 
ionized reflector) for both models (black line) compared to (red) the 
scattered fraction (top panel) and absorbed fraction (bottom panel). 
The flux history of the source reveals how  the 
2000-2001 epochs were dominated by a direct 
view of the power-law continuum 
and associated ionized reflector. The 2005 observation found Mrk 766 
at a very  low flux rising later in the observation. In 2006 we again found 
the source in a fairly 
low state but with a rapid rise evident towards the end of the 
{\it Suzaku} observation.

\begin{figure}
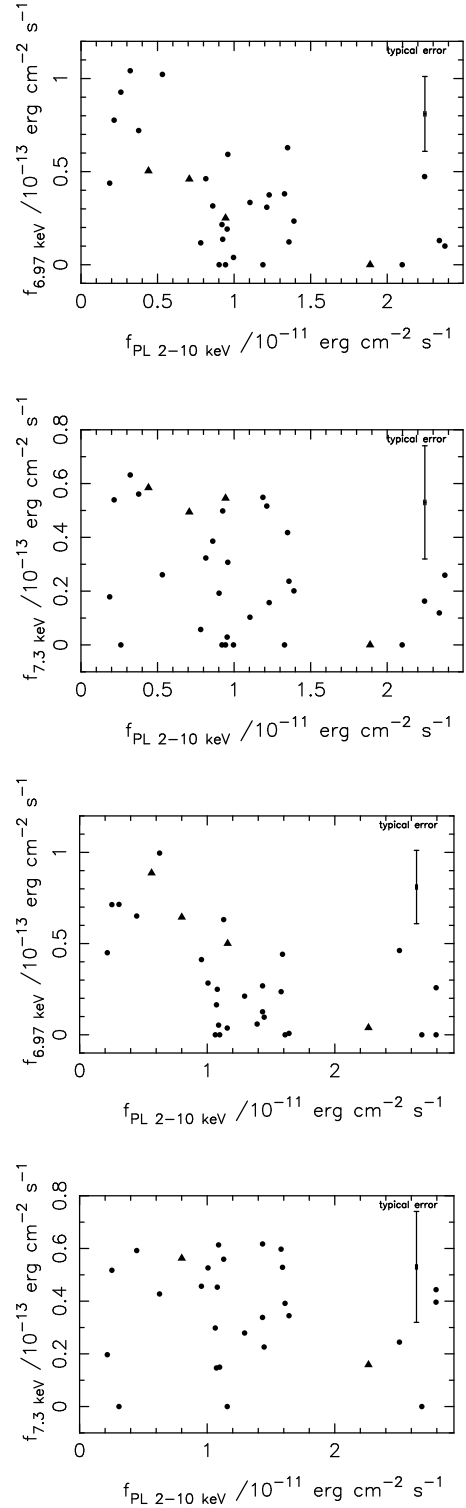

\begin{minipage}{60mm}{
\resizebox{60mm}{!}{
\rotatebox{-90}{
\includegraphics{sc7.ps}
}}}
\end{minipage}

\begin{minipage}{60mm}{
\resizebox{60mm}{!}{
\rotatebox{-90}{
\includegraphics{sc8.ps}
}}}
\end{minipage}

\begin{minipage}{60mm}{
\resizebox{60mm}{!}{
\rotatebox{-90}{
\includegraphics{pc7.ps}
}}}
\end{minipage}

\begin{minipage}{60mm}{
\resizebox{60mm}{!}{
\rotatebox{-90}{
\includegraphics{pc8.ps}
}}}
\end{minipage}
\caption{Absorption line fluxes versus the direct/unabsorbed component  for the scattering model (top two panels) and 
 p-cov model  (bottom two panels) for {\it XMM} (dots) and 
{\it Suzaku} (triangles) data. }
\label{line-directcomp-correlations}
\end{figure}

Overall the  steady level of the scattered flux in Fig.\,\ref{suz_pc_fits} 
is consistent with the PCA, 
that analysis indicated the  scattered component to vary within bounds 
 $\pm 38\%$ of the mean, although the component is higher than expected 
 at the start of the 2000 observations. 
In the partial-covering model, some  anti-correlation of directly-viewed and 
absorbed fractions is a consequence of the model construction.

\begin{table}
\begin{center}
\begin{tabular}{|l|c|c|c|c|}
\hline
          & direct             & hard      & 6.97\,keV & 7.3\,keV \\
\hline
direct    &  -                 &   0.68    & $-0.0013$ & $-0.033$ \\ 
hard      & $-7\times 10^{-7}$ &  -        & 0.50      & $-0.05$  \\
6.97\,keV & $-9\times 10^{-5}$ & 0.0010    &  -        & 0.0061   \\
7.3\,keV  & $-0.1447$          & 0.0949    & 0.0071    &  -       \\
\hline
\end{tabular}
\end{center}\caption{
One-tailed significance level of correlations between parameters in the scattering model
(upper right quadrant) and p-cov absorption model (lower-left quadrant)
fitted to {\it XMM} and {\it Suzaku} 
data  and after allowing column variability for $N_H(1)$.
Anticorrelations are indicated by minus signs in front of each value of significance
level.}
\label{corrtable}
\end{table}

\subsection{Absorption Line Correlations}

Next we  investigate the 
relationship between the absorbed or scattered fraction, and the narrow absorption 
lines. The strength of the correlations between modeled quantities is assessed
using Spearmann's rank correlation coefficient, assuming data points to be
statistically independent, along with correlation parameters for all other combinations of 
variables in both models.  The red-noise power-spectrum 
(see \citealt{miller07}) 
of the intrinsic source variations causes some degree of correlation between
the modeled quantities, so the significance of correlations may be overestimated
by this approach, but without more information on the power-spectrum of 
variations in each quantity it is hard to correct for this effect.  In fact, the
statistical uncertainties on the line flux values are sufficiently large that
this red-noise effect has only a limited influence, although we should bear in
mind that the effect may be present at some level and we should use the
significance values as being a relative measure of which relationships are
the most important rather than being an absolute measurement of significance.

Considering first the scattering model results (Table\,\ref{corrtable}, upper
right quadrant), the flux  of the 6.97 keV 
absorption line appears to be anti- correlated with the directly-viewed component 
and not correlated with the quasi-constant scattered component 
(Figs.\,\ref{line-hardcomp-correlations}, \ref{line-directcomp-correlations}). 

In the p-cov absorption model parameterization the 6.97 keV absorption line is 
anti-correlated with the direct flux or correlated with the 
absorbed flux, with a slight preference for the 
former  scenario 
(Figs.\,\ref{line-hardcomp-correlations}, \ref{line-directcomp-correlations}). 
As the uncovered and covered fractions are anti-correlated in this model, 
it is hard to discern which has the primary correlation with the absorption line. 

The 7.3 keV absorption line flux does not show a correlation with anything in either 
model (Figs.\,\ref{line-hardcomp-correlations},\,\ref{line-directcomp-correlations}). 
However, the  line equivalent width is not consistent with being constant when 
measured against the varying spectral component, but is consistent with a 
constant value against the scattered or absorbed components in the two models.

\section{Discussion}
\label{discussion}

The ambiguity in interpretation of the hard X-ray spectra observed in
the low flux state for many Seyfert galaxies has hindered progress
understanding these sources. We have attempted to move forward by using a 
large dataset accumulated by {\it XMM} over 2000-2005, combined with data 
from the broad bandpass covered by {\it Suzaku} during a 2006 observation, 
for the highly X-ray variable narrow-line Seyfert\,1, Mrk~766. 

Two models have been directly compared in application to time-selected
spectra from Mrk~766.  The aim of the comparison has been 
to distinguish whether large-scale spectral variability is best explained by 
changes in relative contribution from powerlaw plus ionized-reflector 
compared to the 
scattering component, or by variations in the  covering fraction of 
the absorber. It is very important to 
determine the dominant effects as several models that fit the mean 
spectra of Seyfert galaxies comparably well have very different physical interpretations. 

The most robust measurements are those of the directly viewed and scattered or absorbed 
fractions of flux. The obvious interpretations outlined by \citet{miller07} 
are briefly reviewed here to set the scene for further discussion. 
In the scattering model the hard emission component is steady, explained by 
scattering from an extended region such as a disk wind that could also absorb  
scattered photons. PCA limits the flux variations  in this component to 
$\pm 38\%$ of the mean and this relatively low level of variability implies a size of 
at least several light days for this gas but with an indication of 
some response of the inner region of the 
wind to changes in continuum flux. 

 The most likely candidate for the scattering gas is the 
highly-ionized inner zone 
that produces the 6.97 keV absorption line. The scattered fraction of emission 
is a significant fraction of the total and so this zone likely has  a large covering 
fraction, whereas the 
lower ionization layers further out probably have too small a covering to produce 
a significant flux of scattered X-rays. 
 Opacity changes observed in the data could be explained by variations  
in such a wind that affect the degree of self-absorption of the scattered photons. 

Seyfert galaxies generally (e.g. \citealt{vaughanfabian04,pounds04}) show the 
behavior exhibited by Mrk~766, i.e. a hard 
spectral form when the primary continuum flux is faint; this tells us
something about spectral variability in the context of absorption
models. The simple scenario of an intrinsically variable continuum 
with additional variability caused by covering fraction changes from passage of clouds
across the sight-line can be ruled out. The covering fraction would
have to know about the continuum flux for the observed phenomena to be
so widely seen, else we would see some highly absorbed high-state
spectra both in general and also specifically within the long duration Mrk\,766
observations: the observed variability would not be a natural consequence
of such a `mixed variability' picture. Two variants of the p-cov absorption model 
seem plausible however, 
either the continuum drives apparent absorption changes by ionizing some of the gas 
(see later) or the continuum is intrinsically constant and the 
observed variability simply a consequence of the changes in covering fraction.  
The p-cov absorption model fits the {\it XMM} data very well but is a less satisfactory fit to 
the {\it Suzaku} data; as the absorption model was built upon a PCA deconstruction of the {\it XMM} data, 
changes in the absorbing gas between those observations and the {\it Suzaku} epoch would naturally result in 
a poor fit to the latter. 
The more we free the column and ionization-states of the layers, the better the {\it Suzaku} 
data are fitted, so in conclusion, the statistical quality of fits to {\it Suzaku} spectra could be interpreted 
as evidence for absorption changes on year-timescales: these must be expected, when opacity changes on tens of 
ks are also evident in the data.  

One compelling  alternative interpretation of the partial-covering behavior in the 
absorption model 
may be that the region producing the hard X-ray continuum might vary in extent. 
In this alternative picture, when the putative 
coronal region producing X-rays 
is largest the continuum could be visible above gas structures that 
might normally hide the nucleus: this would 
give  apparent changes in covering fraction as  
 the comptonizing region expands and contracts. This could naturally yield a 
constant flux of 
absorbed continuum emission that is obscured and that dominates above 10 keV. 
In this case the correlated 
variations in ionized reflection and power law continuum known in Mrk 766 \citep{miller07} 
could be attributed to  the reflector seeing more continuum 
photons when the coronal region is at its largest. 
However, perhaps a more standard interpretation of the `p-cov absorption' picture
would be to suppose that the absorbing
material is an inhomogeneous disk wind, in which clumps of obscuring material
pass between a central continuum source and the observer.  A wind would naturally have 
the most ionized zones closest to the nucleus, consistent with these data. 

Considering the second-order spectral  variability  
(not addressed in the principal components analysis of \citealt{miller07}): 
we find  opacity variations in one of the absorption layers can account for most of the observed 
 second-order effects. 
Degeneracy between column density and ionisation parameter means we cannot isolate the 
origin of the opacity changes with current data but again, the opacity variations 
support the general picture of an important disk wind component, driving the 
observed spectral variability.  

It is interesting to note that both models presented here are broadly consistent with the 
soft-band data down to 0.5 keV. The partial-covering absorption model provides a particularly good fit and 
naturally explains the soft-band spectral curvature without recourse to any separate ``soft excess'' component. 
As the partial-covering model was not fit down to 0.5 keV, then the agreement of soft-band data with the 
extrapolation of the partial-covering model to soft energies, supports the applicability of that model.

One important diagnostic result is that the absorption line at 6.97 keV,
discovered in the mean spectrum \citep{miller07} shows significant
variability, correlated with other aspects of source behavior.
Interestingly the absorption line itself is not modeled within any of
the zones of gas used to describe the overall spectral curvature but, as evident in Tables 
A1, A2, the opacity changes themselves are anti-correlated with the flux of the direct component. 
All zones of gas may be responding to continuum changes, or continuum changes themselves 
may be  an artifact of variable absorption. 
This possibility has been considered previously: 
the tight correlation between the ionized reflector and continuum  flux  
might  be a consequence of variable occultation of both if 
Compton-thick `bricks' 
of variable  covering fraction  occult continuum and reprocessed emission from within. 
The bricks would be implied to exist  $\sim 70 r_g$ 
\citep{miller07} which is close to where the occulted line is estimated to arise 
(based on the emission line width). By similar arguments, opacity changes observed on 
comparable timescales driven by motion of clouds 
across the sight-line would indicate clouds at the same radial location. If the opacity changes 
originate as ionisation-state variations then the location is currently unconstrained. 

The existence of a strong isolated absorption line at 6.97 keV 
yields the lower limit  $\log\xi \ga 4$, ie. it comes from gas that is the most 
highly-ionized of all layers 
evident in the 2-10 keV band. It is clear that the other layers of lower $\xi$ gas may exist along 
the same sight-line but at larger radial distances from the central source: the 
highest-$\xi$ zone could not exist outside the other gas layers as shielding 
would prevent it achieving  $\log\xi \ga 4$. Similar models have been discussed 
 for sources such as e.g., 
NGC 3516 \citep{turnerea05}  
and NGC 4151 \citep{kraemerea4151} and thus it seems that spectral variations driven by 
a time-variable disk wind may be applicable across the Seyfert population. 

The 6.97 keV absorption line  is so 
strong that it is not possible to model the zone from which it
originates without invoking a significant turbulent velocity for the
absorbing gas, again, consistent with its origin very close to the active nucleus. 
Low turbulent velocities (few hundred
km/s) predict much weaker lines even for columns densities of 
few $\times 10^{24} {\rm cm^{-2}}$ as  the line saturates quickly and the 
equivalent width predicted when line 
broadening occurs can be much higher. 
One must invoke a turbulent velocity
$\ga 2000$\,km\,s$^{-1}$ to fit the observed line at its
deepest with a column that is Compton-thin; for a turbulent velocity
of 3000\,km\,s$^{-1}$ we find a fit for the line in the mean spectrum 
using a column of gas with 
 $N_H \sim 5 \times 10^{23} {\rm cm^{-2}}$. 
Variability in the absorption line at 6.97 keV supports the general conclusion 
that rapid absorption variation is 
very important: the presence of such a line strengthens previous suggestions 
\citep{np94,turnerea05,reevesea05}  that 
very large column densities of highly-ionized gas exist close to the nucleus, dangerously close to 
the conditions that cause most confusion with Fe emission features. 

\begin{figure}
\begin{minipage}{80mm}{
\resizebox{80mm}{!}{
\rotatebox{-90}{
\includegraphics{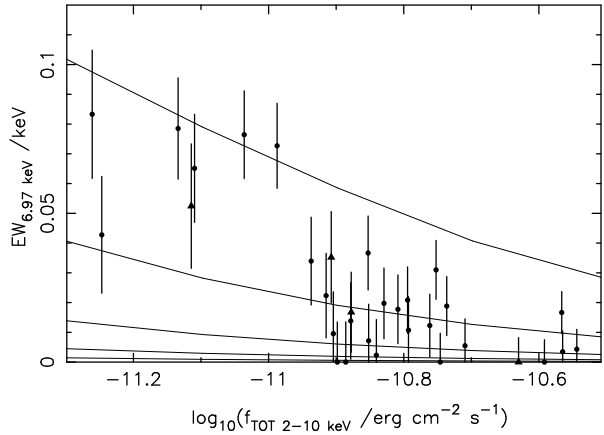}}}}
\end{minipage}
\caption
{Variation of equivalent width of the 6.97 keV line, measured 
against the total source continuum, with the total source flux (points 
with error bars: for {\it XMM} (dots) and 
{\it Suzaku} (traingles) data.). 
Values are calculated within the scattering model. Also 
shown are model lines for the variation expected from {\sc XSTAR} models 
of the Fe {\sc XXVI} 6.97\,keV line arising from ionisation variation alone, 
for column densities with solar abundance of $N_H = 10^{21.5}, 10^{22}, 
10^{22.5}, 10^{23}, 10^{23.5} $\,cm$^{-2}$.
}
\label{EW}
\end{figure}

The absorption-line correlations found here seem to present two possibilities:
in the p-cov 
absorption  model: the 6.97\,keV line flux variations could be due either to 
being constant EW absorption in front of the fluctuating absorbed part, or to
being over-ionized at times of exposure to high powerlaw flux.  
Fig.\,\ref{line-hardcomp-correlations} shows that the 6.97\,keV 
line flux does indeed correlate positively with the flux in the 
absorbed component, lending weight to the first of these possibilities.
There is however a stronger anti-correlation with the direct component
(Fig.\,\ref{line-directcomp-correlations}, Table\,\ref{corrtable}),
which appears to show a sudden transition in line flux for direct
component flux $f_{\rm 2-10 keV} \ga 10^{-11}$\,erg\,cm$^{-2}$\,s$^{-1}$.
We discuss below the difficulty of obtaining such a rapid transition
purely from changing ionization parameter.

Compared with the 6.97\,keV line variations, 
the 7.3\,keV line behavior is less clear due to its model-dependency.   
One thing that is clear is that the feature is persistent, 
remaining evident in {\it Suzaku} data  
during 2006 and that it is not an artifact of any instrument.  However, the 
identification of this line 
is unknown, there are no known transitions expected 
of this strength at such a rest-energy without a substantial complex of other
lines appearing \citep{kallman04}.
It is possible that this feature represents an ionized Fe edge whose shape is 
distorted due to unresolved features close in energy, or as noted by \citet{miller07} the feature 
could arise in a high-velocity outflow such as from Fe\,{\sc XXVI}\,Ly$\alpha$ in gas with a bulk 
outflow velocity 
$\sim 13,000$\,km$^{-1}$.  
There is a suggestion that the 7.3 keV line variations are correlated with
those of the 6.97\,keV line, however,  there is no evidence for a causal relation
between the 7.3\,keV line and  the direct component 
in the p-cov absorption model, and only weak evidence for correlation
with the direct component in the scattering model 
(Table\,\ref{corrtable}). This lack of correlation is, in itself interesting.
The equivalent width of the 7.3 keV line is inconsistent  with a constant value when 
measured against the variable spectral component in either model, but is consistent 
with a constant value when measured against the hard/steady component in either model: this 
 may be associated with a quasi-constant continuum component, leaving the equivalent 
width variations 
attributed to changes in the flux of the direct continuum component. 

In the scattering model, the variation in 6.97\,keV line flux would need to
be explained as arising from a change in ionization state of the 
absorbing gas as the source brightens overall.  It appears that the
decrease in line flux is associated with quite a sharp transition in
total source brightness which would not be expected given the broad
range of ionization parameter over which any given ionization state
generally exists \citep{kallman04}.  We can test this explanation
by generating model absorbers with {\sc xstar} assuming that the line
is Fe\,XXVI\,6.97\,keV and assuming that the ionization parameter
varies with the total source brightness.  
As discussed above, the apparent lack of Fe\,K$\alpha$\,6.7\,keV 
in absorption indicates high ionization, $\log\xi>4$. 
As also discussed above, in order to obtain a high equivalent width
without an extreme column density some velocity broadening is probably
required, and we test models with velocity broadening
$\sigma =$ 3000\,km\,s$^{-1}$, a value that allows reconciliation of the 
measured line strength with physical models, while staying within the 
velocity constraint yielded from the data ($\sigma < 6000$\,km\,s$^{-1}$). 
The results are shown in Fig\,\ref{EW}, where the equivalent width has
been measured against the total continuum flux: i.e. it has been assumed that
the absorbing gas lies in front of all the source emission components.
The radial distance and density of the absorbing material is unknown, so we 
have investigated the cases in the range $n=10^{8} - 10^{12}$\,cm$^{-3}$ and 
choose to adjust the ionization parameter so that the maximum possible
variation in line equivalent width is produced, corresponding to $\log\xi \sim 5$.

It appears that indeed the rapid variation in equivalent width cannot be well
reproduced simply by changes in source brightness affecting the ionization
state of the absorber.  It may be, however, that the particular  absorbing zone lies only 
in front of the scattered component of emission, but that nonetheless its
ionization state is still determined by the total source brightness.  In this
case the typical equivalent width measured against the scattered component
is higher, but shows less variation.  Although the ionization-variation
hypothesis still does not produce good agreement with the observed variation, 
such a picture probably cannot be ruled out. This deduction does have 
an important implication for the nature of the scattering component.
If the 6.97\,keV absorbing gas has to be associated with this hard spectral component only, then
that component 
cannot be significantly relativistically blurred, otherwise the absorption
line would also be blurred.  But if the absorbing gas lies outside any such
blurred region, the rapid variation in EW with source flux would not take place.
This consideration seems to rule out relativistically blurred models for the
scattered component, unless an alternative blurred model can be found that reduces
the significance of the 6.97\,keV absorption line.

\section{Conclusions} 

Detailed time-resolved X-ray spectroscopy has  
confirmed that the spectral shape of Mrk~766 is directly coupled 
to the source brightness. Two model interpretations of principal components analysis 
fit the time-sliced data well, with 
changes in spectral shape then attributed to the relative strength of the continuum 
and a self-absorbed scattered component, or to changes in covering fraction of an absorber. 
In addition to this dominant mode of variation, significant temporal 
variations in opacity down to timescales of tens of ks are also required to fit the time-sliced data, as 
previously indicated from residuals in the principal components analysis \citep{miller07}. 
The previously-reported  absorption line at  6.97 keV is likely Fe {\sc XXVI} K$\alpha$  from gas 
with $\log\xi\ga4$ and this varies correlated with the primary continuum flux,  
tracing the behavior of a highly-ionized gas zone and extending our understanding of 
the parameter-space spanned by the complex, multi-layered absorber in this Seyfert galaxy. 

In conclusion then, Mrk 766 shows  strong evidence for  stratified layers of absorption, possibly 
decreasing in ionization 
with increasing radial 
location. The gas layers have rapidly  variable covering factor, opacity and ionization 
state and are required to explain the spectral variability in this source, whether 
or not one allows the lowest 
flux state to be dominated by some residual scattered continuum component. Some scattered and 
reflected emission must be present to explain the observations, especially the rapidly variable
emission features 
correlation with continuum flux, that show up in the high state. 
These results,   combined with evidence in the literature 
for  significant 
bulk outflow velocities in X-ray absorbers in general 
\citep{crenshaw03}  points to  
disk winds as dominant in shaping Seyfert spectral variations in the 
X-ray band.

\section{Acknowledgements}
We thank Brian Wingert for help with the {\it Suzaku} data reduction. 
We thanks the referee, Chris Done, for comments that helped us 
clarify and otherwise improve the paper.

\appendix

\newpage

\section{Model fit parameters}

\begin{table*}[!h]
\begin{centering}
\begin{tabular}{|lcccccccc|} 	
\hline 
\multicolumn{9}{|c|} 
	{\rule[-3mm]{0mm}{8mm}Scattering Model} \\
\hline 
\hline 
 & & & & & & & & \\
Time & Total & Direct & Scattered & 6.97 keV & 7.30 keV & 6.4 keV & $N_H(1)$ & $\chi^2$ $^{1}$ \\
Slice & \multicolumn{3}{c}{2-10 keV flux /$10^{-11}$ erg cm$^{-2}s^{-1}$} & \multicolumn{3}{c}{Line photon flux /$10^{-6}$ cm$^{-2}s^{-1}$} & /$10^{22}$cm$^{-2}$ & \\
\multicolumn{9}{|c|}{2000 {\it XMM}}\\
  1 & 1.406 & 0.782$\pm$0.010 & 0.510$\pm$0.040 &  1.1$\pm$2.0 &  0.5$\pm$1.9 &  9.4$\pm$1.0 &  0.8$^{+0.2}_{-0.1}$ & 1.00\\
  2 & 1.611 & 0.955$\pm$0.024 & 0.604$\pm$0.050 &  1.7$\pm$2.9 &  0.3$\pm$2.9 &  4.5$\pm$2.0 &  0.4$^{+0.1}_{-0.1}$ & 1.03\\
\multicolumn{9}{|c|}{2001 {\it XMM}}\\
  3 & 2.724 & 2.377$\pm$0.006 & 0.296$\pm$0.036 &  0.9$\pm$2.9 &  2.2$\pm$2.8 &  4.5$\pm$3.0 &  8.3$^{+0.7}_{-0.5}$ & 1.56\\
  4 & 2.560 & 2.098$\pm$0.007 & 0.344$\pm$0.030 &  0.0$\pm$2.7 &  0.0$\pm$2.7 &  9.1$\pm$2.0 &  3.3$^{+0.7}_{-0.7}$ & 1.29\\
  5 & 2.715 & 2.244$\pm$0.007 & 0.401$\pm$0.030 &  4.2$\pm$2.7 &  1.4$\pm$2.7 &  6.4$\pm$2.0 &  3.6$^{+0.6}_{-0.8}$ & 1.28 \\
  6 & 2.858 & 2.340$\pm$0.004 & 0.402$\pm$0.030 &  1.2$\pm$2.9 &  1.0$\pm$2.8 &  9.7$\pm$2.0 &  3.6$^{+0.6}_{-0.7}$ & 1.33\\
  7 & 1.835 & 1.329$\pm$0.007 & 0.433$\pm$0.026 &  3.4$\pm$2.5 &  0.0$\pm$2.7 &  6.2$\pm$2.0 &  2.3$^{+0.4}_{-0.3}$ & 1.01\\
\multicolumn{9}{|c|}{2005 {\it XMM}} \\
  8 & 0.548 & 0.216$\pm$0.003 & 0.318$\pm$0.017 &  7.0$\pm$1.3 &  4.6$\pm$1.2 &  2.3$\pm$2.0 &  7.9$^{+0.3}_{-0.3}$ & 1.20\\
  9 & 0.920 & 0.322$\pm$0.003 & 0.555$\pm$0.017 &  9.3$\pm$1.6 &  5.4$\pm$1.5 &  5.0$\pm$2.0 &  3.5$^{+0.2}_{-0.2}$ & 1.37\\
 10 & 0.735 & 0.260$\pm$0.003 & 0.431$\pm$0.018 &  8.3$\pm$1.4 &  0.0$\pm$1.6 &  4.3$\pm$3.0 &  5.8$^{+0.2}_{-0.2}$ & 0.87\\
 11 & 0.566 & 0.187$\pm$0.004 & 0.368$\pm$0.027 &  3.9$\pm$2.4 &  1.5$\pm$2.4 &  1.5$\pm$3.0 &  9.2$^{+0.1}_{-1.0}$ & 1.16\\
 12 & 0.777 & 0.377$\pm$0.004 & 0.355$\pm$0.018 &  6.5$\pm$1.5 &  4.8$\pm$1.4 &  4.8$\pm$3.0 &  4.7$^{+0.3}_{-0.3}$ & 1.12\\
 13 & 1.404 & 0.959$\pm$0.005 & 0.349$\pm$0.021 &  5.3$\pm$1.9 &  2.6$\pm$1.8 &  8.6$\pm$3.0 &  2.4$^{+0.4}_{-0.3}$ & 1.26\\
 14 & 1.217 & 0.861$\pm$0.004 & 0.297$\pm$0.020 &  2.8$\pm$1.8 &  3.3$\pm$1.8 &  5.5$\pm$3.0 &  3.7$^{+0.5}_{-0.1}$ & 1.20\\
 15 & 1.156 & 0.815$\pm$0.004 & 0.295$\pm$0.023 &  4.1$\pm$1.9 &  2.8$\pm$1.8 &  4.5$\pm$2.0 &  4.2$^{+0.5}_{-0.6}$ & 1.10\\
 16 & 1.301 & 0.944$\pm$0.005 & 0.307$\pm$0.021 &  0.0$\pm$1.9 &  0.0$\pm$1.9 &  4.1$\pm$2.0 &  4.0$^{+0.5}_{-0.5}$ & 1.38\\
 17 & 1.554 & 1.213$\pm$0.005 & 0.299$\pm$0.025 &  2.8$\pm$2.0 &  4.4$\pm$1.9 &  4.1$\pm$3.0 &  4.9$^{+0.6}_{-0.6}$ & 1.31\\
 18 & 1.444 & 0.996$\pm$0.005 & 0.418$\pm$0.022 &  0.4$\pm$2.1 &  0.0$\pm$2.1 &  2.0$\pm$3.0 &  3.8$^{+0.4}_{-0.4}$ & 1.10\\
 19 & 1.264 & 0.902$\pm$0.004 & 0.337$\pm$0.025 &  0.0$\pm$2.2 &  1.6$\pm$2.1 &  2.2$\pm$2.0 &  4.9$^{+0.5}_{-0.5}$ & 1.11\\
 20 & 1.482 & 1.104$\pm$0.004 & 0.345$\pm$0.023 &  3.0$\pm$2.0 &  0.9$\pm$1.9 &  3.1$\pm$3.0 &  4.5$^{+0.5}_{-0.5}$ & 1.15\\
 21 & 1.732 & 1.391$\pm$0.004 & 0.307$\pm$0.026 &  2.1$\pm$2.1 &  1.7$\pm$2.0 &  3.2$\pm$3.0 &  5.3$^{+0.6}_{-0.6}$ & 1.04\\
 22 & 1.953 & 1.358$\pm$0.004 & 0.566$\pm$0.024 &  1.1$\pm$2.4 &  2.0$\pm$2.3 &  2.7$\pm$2.0 &  2.9$^{+0.2}_{-0.3}$ & 1.20\\
 23 & 1.795 & 1.188$\pm$0.004 & 0.543$\pm$0.028 &  0.0$\pm$2.7 &  4.7$\pm$2.4 &  5.7$\pm$3.0 &  2.4$^{+0.3}_{-0.3}$ & 1.18\\
 24 & 1.769 & 1.349$\pm$0.004 & 0.428$\pm$0.026 &  5.6$\pm$2.2 &  3.6$\pm$2.1 &  0.3$\pm$3.0 &  5.1$^{+0.4}_{-0.5}$ & 1.06\\
 25 & 1.606 & 1.229$\pm$0.004 & 0.366$\pm$0.024 &  3.4$\pm$2.0 &  1.3$\pm$2.0 &  1.4$\pm$3.0 &  4.4$^{+0.5}_{-0.5}$ & 1.20\\
 26 & 1.247 & 0.925$\pm$0.004 & 0.284$\pm$0.023 &  1.22$\pm$1.9 &  4.3$\pm$1.8 &  3.6$\pm$2.0 &  4.4$^{+0.6}_{-0.6}$ & 0.96\\
 27 & 1.030 & 0.532$\pm$0.003 & 0.399$\pm$0.030 &  9.2$\pm$2.6 &  2.2$\pm$2.7 &  0.9$\pm$2.0 &  4.3$^{+0.5}_{-0.5}$ & 0.82\\
 28 & 1.323 & 0.921$\pm$0.002 & 0.365$\pm$0.020 &  1.9$\pm$1.9 &  0.0$\pm$1.9 &  3.2$\pm$2.0 &  4.3$^{+0.4}_{-0.4}$ & 0.87\\
\multicolumn{9}{|c|}{2006 {\it Suzaku}}\\
 29 & 1.325 & 0.944$\pm$0.009 & 0.297$\pm$0.020 &  2.2$\pm$2.4 &  4.7$\pm$2.4 &  6.3$\pm$1.7 &  2.9$^{+0.1}_{-0.1}$ & 1.17\\
 30 & 0.768 & 0.441$\pm$0.004 & 0.215$\pm$0.016 &  4.5$\pm$1.8 &  5.0$\pm$1.8 &  8.4$\pm$1.3 &  2.9$^{+0.9}_{-0.6}$ & 0.99\\
 31$^2$ & 1.238 & 0.707$\pm$0.009 & 0.407$\pm$0.020 &  4.1$\pm$1.8 &  2.8$\pm$1.8 &  1.0$\pm$1.3 &  1.3$^{+0.3}_{-0.3}$ & 1.16\\
 32 & 2.344 & 1.888$\pm$0.020 & 0.443$\pm$0.050 &  0.0$\pm$1.8 &  0.0$\pm$8.0 &  1.5$\pm$8.8 &  2.7$^{+1.3}_{-0.8}$ & 0.71\\
\hline
\end{tabular}  
\caption{
Values of fitted parameters for the `scattering' model, in each time slice,
as described in the text.  {\em Notes:}
$(1)$ XMM fits have 133 degrees of freedom, Suzaku fits have 145 degrees of freedom. 
$(2)$ Slice 31 required $N_H(2)$  free, with value
3.4$^{+1.1}_{-1.5} \times 10^{22}{\rm cm^{-2}}$.
}
\end{centering}
\end{table*}
Tables A1 and A2 give the model fit parameters for the scattering and absorption 
models applied to the {\it XMM} and {\it Suzaku} data.  The model components are
described in the main body of the paper.  
For each table, the values given in each
time slice (column one) are: the $2-10$\,keV fluxes of the total, direct and scattered
or absorbed components (columns $2-4$); the absorption-line flux in each of the 6.97\,keV
and 7.3\,keV (rest-frame) absorption features (columns 5,6); the emission-line flux in
the 6.4\,keV emission line (column 7); the $N_H$ column density (column 8); and the
reduced $\chi_r^2$, noting that the {\em XMM} and Suzaku fits have differing degrees
of freedom.

\begin{table*}
\begin{centering}
\begin{tabular}{|lcccccccc|} 	
\hline 
\multicolumn{9}{|c|} 
	{\rule[-3mm]{0mm}{8mm}Partial Covering Model} \\
\hline 
\hline 
 & & & & & & & & \\
Time  & Total & Direct & Absorbed & 6.97 keV & 7.30 keV & 6.4 keV & $N_H(1)$ & $\chi_r^2$ $^1$\\
Slice & \multicolumn{3}{c}{2-10 keV flux /$10^{-11}$ erg cm$^{-2}s^{-1}$} & \multicolumn{3}{c}{Line photon flux /$10^{-6}$ cm$^{-2}s^{-1}$} & /$10^{22}$cm$^{-2}$ & \\
\multicolumn{9}{|c|}{2000 {\it XMM}}\\
  1 & 1.422 & 1.079$\pm$0.010 & 0.348$\pm$0.040 &  2.2$\pm$2.0 &  3.9$\pm$1.8 &  0.0$\pm$5.0 &  0.0$^{+0.5}_{-0.0}$ & 1.04\\
  2 & 1.617 & 1.446$\pm$0.024 & 0.173$\pm$0.050 &  0.9$\pm$2.9 &  1.9$\pm$2.7 &  0.0$\pm$5.0 &  0.4$^{+0.4}_{-0.4}$ & 1.13\\
\multicolumn{9}{|c|}{2001 {\it XMM}}\\
  3 & 2.797 & 2.793$\pm$0.006 & 0.000$\pm$0.036 &  0.0$\pm$2.8 &  3.4$\pm$2.5 & 4.5$\pm$5.2 &  4.0$^{+100.0}_{-4.0}$ & 0.87\\
  4 & 2.556 & 2.507$\pm$0.007 & 0.035$\pm$0.030 &  4.1$\pm$2.7 &  2.1$\pm$2.5 & 7.8$\pm$5.1 &  0.69$^{+11.3}_{-0.7}$ & 0.87\\
  5 & 2.715 & 2.682$\pm$0.007 & 0.035$\pm$0.030 &  0.0$\pm$2.7 &  0.0$\pm$2.5 & 4.6$\pm$5.0 &  0.28$^{+8.8}_{-0.7}$ & 0.79 \\
  6 & 2.859 & 2.794$\pm$0.004 & 0.061$\pm$0.030 &  2.3$\pm$2.9 &  3.8$\pm$2.6 & 7.2$\pm$5.0 &  0.36$^{+2.7}_{-0.4}$ & 0.83\\
  7 & 1.843 & 1.610$\pm$0.007 & 0.238$\pm$0.026 &  0.0$\pm$2.5 &  3.4$\pm$2.5 &  0.0$\pm$5.0 &  0.36$^{+2.8}_{-0.4}$ & 0.83\\
\multicolumn{9}{|c|}{2005 {\it XMM}} \\
  8 & 0.550 & 0.253$\pm$0.003 & 0.307$\pm$0.017 &  6.4$\pm$1.3 &  4.4$\pm$1.1 &  2.4$\pm$2.9 &  9.34$^{+1.7}_{-1.3}$ & 1.01\\
  9 & 0.930 & 0.366$\pm$0.003 & 0.578$\pm$0.017 & 11.0$\pm$1.6 &  9.4$\pm$1.4 &  3.5$\pm$3.5 &  1.47$^{+0.7}_{-0.3}$ & 1.29\\
 10 & 0.738 & 0.309$\pm$0.003 & 0.435$\pm$0.018 &  6.4$\pm$1.4 &  0.0$\pm$1.5 &  3.0$\pm$3.2 &  5.39$^{+0.0}_{-0.0}$ & 0.80\\
 11 & 0.572 & 0.217$\pm$0.004 & 0.361$\pm$0.027 &  4.0$\pm$2.4 &  1.7$\pm$2.2 &  0.9$\pm$4.4 & 12.28$^{+0.1}_{-0.1}$ & 1.07\\
 12 & 0.779 & 0.447$\pm$0.004 & 0.339$\pm$0.018 &  5.8$\pm$1.5 &  5.1$\pm$1.4 & 4.8$\pm$3.3   &  3.85$^{+0.6}_{-0.5}$ & 1.11\\
 13 & 1.405 & 1.129$\pm$0.005 & 0.280$\pm$0.021 &  5.7$\pm$1.9 &  4.8$\pm$1.7 & 6.2$\pm$4.0 &  0.05$^{+1.1}_{-0.1}$ & 1.26\\
 14 & 1.218 & 1.006$\pm$0.004 & 0.213$\pm$0.020 &  2.5$\pm$1.8 &  4.5$\pm$1.6 & 4.8$\pm$3.7 &  0.65$^{+1.7}_{-0.7}$ & 1.08\\
 15 & 1.159 & 0.955$\pm$0.004 & 0.206$\pm$0.023 &  3.7$\pm$1.9 &  3.9$\pm$1.7 & 5.2$\pm$4.0 &  1.25$^{+3.8}_{-0.5}$ & 0.93\\
 16 & 1.303 & 1.097$\pm$0.005 & 0.202$\pm$0.021 &  0.0$\pm$1.9 &  1.3$\pm$1.7 &  3.0$\pm$3.7 &  0.16$^{+2.9}_{-0.2}$ & 1.01 \\
 17 & 1.552 & 1.434$\pm$0.005 & 0.122$\pm$0.025 &  1.1$\pm$2.0 &  5.3$\pm$1.8 &  3.6$\pm$3.7 &  3.50$^{+2.6}_{-3.0}$ & 0.97 \\
 18 & 1.450 & 1.157$\pm$0.005 & 0.289$\pm$0.022 &  0.3$\pm$2.1 &  0.0$\pm$2.0 &  0.0$\pm$4.0 &  0.24$^{+0.7}_{-0.2}$ & 0.77\\
 19 & 1.266 & 1.063$\pm$0.004 & 0.202$\pm$0.025 &  0.0$\pm$2.2 &  2.6$\pm$1.9 &  1.2$\pm$5.0 &  3.26$^{+1.5}_{-1.5}$ & 0.93\\
 20 & 1.477 & 1.293$\pm$0.004 & 0.187$\pm$0.023 &  1.9$\pm$2.0 &  2.4$\pm$1.8 &  1.2$\pm$4.0 &  1.36$^{+2.6}_{-1.4}$ & 0.93\\
 21 & 1.731 & 1.641$\pm$0.004 & 0.092$\pm$0.026 &  0.1$\pm$2.1 &  3.0$\pm$1.9 &  1.4$\pm$3.8 &  4.85$^{+9.1}_{-4.8}$ & 0.79\\
 22 & 1.960 & 1.579$\pm$0.004 & 0.386$\pm$0.024 &  2.1$\pm$2.4 &  5.1$\pm$2.1 &  1.2$\pm$4.0 &  0.05$^{+0.6}_{-0.0}$ & 1.06\\
 23 & 1.797 & 1.391$\pm$0.004 & 0.410$\pm$0.028 &  0.5$\pm$2.7 &  7.8$\pm$2.3 &  2.9$\pm$4.0 &  0.05$^{+0.5}_{-0.0}$ & 1.12\\
 24 & 1.774 & 1.590$\pm$0.004 & 0.193$\pm$0.026 &  4.0$\pm$2.1 &  4.5$\pm$2.0 &  0.0$\pm$5.0 &  3.16$^{+1.3}_{-2.0}$ & 0.78\\
 25 & 1.609 & 1.434$\pm$0.004 & 0.180$\pm$0.024 &  2.4$\pm$2.0 &  2.9$\pm$1.9 &  0.0$\pm$5.0 &  0.05$^{+2.0}_{-0.0}$ & 0.94\\
 26 & 1.249 & 1.089$\pm$0.004 & 0.162$\pm$0.023 &  0.5$\pm$1.9 &  5.3$\pm$1.7 &  3.1$\pm$5.0 &  1.84$^{+2.2}_{-1.8}$ & 0.69\\
 27 & 1.031 & 0.626$\pm$0.003 & 0.408$\pm$0.030 &  8.9$\pm$2.6 &  3.7$\pm$2.5 & 8.3$\pm$5.0 &  2.82$^{+0.6}_{-0.3}$ & 0.82\\
 28 & 1.325 & 1.073$\pm$0.002 & 0.252$\pm$0.020 &  1.5$\pm$1.9 &  1.3$\pm$1.8 &  1.6$\pm$3.6 &  0.83$^{+3.0}_{-0.8}$ & 0.74\\
\multicolumn{9}{|c|}{2006 {\it Suzaku}}\\
 29 & 1.316 & 1.160$\pm$0.009 & 0.167$\pm$0.020 &  4.5$\pm$2.5 &  7.9$\pm$2.2 &  2.5$\pm$5.2 &  0.0$^{+0.9}_{-0.6}$ &  1.38 \\
 30 & 0.764 & 0.564$\pm$0.004 & 0.215$\pm$0.016 &  7.9$\pm$1.9 &  7.9$\pm$1.7 &  2.8$\pm$4.1 &  0.0$^{+0.3}_{-0.0}$ & 1.40 \\
 31$^2$ & 1.238 & 0.801$\pm$0.009 & 0.444$\pm$0.020 & 5.8$\pm$2.3 &  4.7$\pm$2.1 &  0.0$\pm$4.9 &  5.0$^{+0.6}_{-0.7}$ & 1.14\\
 32 & 2.339 & 2.264$\pm$0.020 & 0.077$\pm$0.050 &  0.4$\pm$8.1 &  1.4$\pm$8.0 &  0.0$\pm$14.0 &  0.0$^{+4.9}_{-0.0}$ & 0.69\\
\hline
\end{tabular} 
\caption{
Values of fitted parameters for the `absorption' model, in each time slice,
as described in the text.  {\em Notes:}
$(1)$ XMM fits have 133 degrees of freedom, Suzaku fits have 145 degrees of freedom. 
$(2)$ Slice 31 required $N_H(2)$  free, with value
$0.0 < 0.5 \times 10^{22}{\rm cm^{-2}}$.
} 
\end{centering}
\end{table*}

\label{lastpage}


\begin{thebibliography}{99}
\setlength\itemsep{0mm}
\bibitem[Arnaud(1996)]{arnaud}
         Arnaud, K., 1996, 
Astronomical Data Analysis Software and Systems V, A.S.P. Conference Series, 
ed. G.H. Jacoby \& J. Barnes, 101, 17
\bibitem[Boldt \&   Leiter(1987)]{boldt87}
         Boldt, E., Leiter, D., 1987, \apj 322, 1 
\bibitem[Crenshaw, Kraemer \& George(2003)]{crenshaw03}
        Crenshaw, D.M., Kraemer, S.B. \& George, I.M. 2003, ARA\&A,  41, 117 
\bibitem[George \& Fabian(1991)]{fgf}
         George, I.M. \& Fabian, A.C. 1991, \mnras, 249, 352\bibitem[Gruber et al(1999)]{gruber99}
        Gruber, D.E., Matteson, J.L., Peterson, L.E., Jung, G.V., 1999, \apj 520, 124 
\bibitem[Guilbert \& Rees(1988)]{guilbertrees88}
         Guilbert, P.W. \& Rees, M.J. 1988, \mnras, 233, 475 
\bibitem[Kallman et al.(2004)]{kallman04}
         Kallman, T., Palmeri, P., Bautista, M.A., Mendoza, C. \& Krolik, J.H.
         2004, \apjs, 155, 675
\bibitem[Kraemer et al(2005)]{kraemerea4151}
        Kraemer, S.B., et al 2005, \apj, 633, 693 
\bibitem[Laor et al.(1991)]{laor91}
         Laor, A. 1991, \apj, 376, 90
\bibitem[Lightman \& White(1988)]{lightmanwhite88}
  Lightman, A.P. \& White,T.R. 1988, \apj, 335, 57
\bibitem[Miller et al.(2006)]{miller06}
         Miller, L., Turner, T.J., Reeves, J.N. et al. 2006, A\&A, 453, L13
\bibitem[Miller et al.(2007)]{miller07}
         Miller, L., Turner, T.J., Reeves, J.N. et al. 2007, A\&A, 463, 131 
\bibitem[Nandra \& Pounds(1994)]{np94}
         Nandra, K., Pounds, K.A. 1994, \mnras, 268, 405 
\bibitem[Nandra et al.(1997)]{nandraea97}
         Nandra, K., George, I.M., Mushotzky, R.F., Turner, T.J. \& Yaqoob, T. 
         1997, \apj 477, 602 
\bibitem[Ogle et al(2003)]{og03}
         Ogle, P.M., Brookings, T., Canizare, C.R., Lee, J.C., Marshall, H.L., A \& A, 402, 849  
\bibitem[Osterbrock \& Pogge(1985)]{osterbrock}
         Osterbrock, D.E. \& Pogge, R.W. 1985, \apj, 297, 166
\bibitem[Perola et al.(2002)]{perola02}
         Perola, G.C., Matt, G., Cappi, M., Fiore, F., Guainazzi, M., Maraschi, L., Petrucci, P.O. 
         \& Piro, L. 2002, A\&A 389, 802  
\bibitem[Pounds et al(2004)]{pounds04}
        Pounds, K.A., Reeves, J.N., Page, K.L., O'Brien, P.T., 2004, \apj 605, 670 
\bibitem[Reeves et al.(2004)]{reevesea05}
         Reeves, J.N., Nandra, K., George, I.M., Pounds, K.A., Turner, T.J. \&
         Yaqoob, T. 2004, \apj, 602, 648
\bibitem[Ross \& Fabian(2005)]{rossfabian}
         Ross, R.R. \& Fabian, A.C. 2005, \mnras, 358, 211 
\bibitem[Str\"{u}der et al.(2001)]{struder}
  Str\"{u}der, L., Briel, U., Dennerl, K. et al. 2001, A\&A, 365, L18
\bibitem[Tanaka et al.(1995)]{tanaka95} 
         Tanaka, Y., Nandra, K., Fabian, A. et al. 1995, Nature, 375, 659 
\bibitem[Turner et al.(2005)]{turnerea05}
         Turner, T.J., Kraemer, S.B., George, I.M., Reeves, J.N. \& Bottorff, M.C.
          2005, \apj, 618, 155
\bibitem[Turner et al.(2006)]{turner766}
         Turner, T.J., Miller, L., George, I.M. \& Reeves, J.N. 2006, A\&A, 445, 59
\bibitem[Vaughan \& Fabian(2004)]{vaughanfabian04}
         Vaughan, S. \& Fabian, A.C. 2004, \mnras, 348, 1415 
\bibitem[Zdziarski et al.(1995)]{zdziarski95}
         Zdziarski, A.A., Johnson, W.N., Done, C., Smith, D., 
         \& McNaron-Brown, K. 1995, \apj 438, L63 
\end{thebibliography}
\end{document}